\documentclass[a4paper,11pt]{article}

\usepackage[UKenglish]{babel}
\usepackage[margin=2.5cm]{geometry}
\usepackage[hypertexnames=false]{hyperref}
\usepackage[noend]{algpseudocode}
\usepackage{algorithm}
\usepackage{amsmath,amssymb,mathtools}
\usepackage{cleveref}
\usepackage{graphicx}
\usepackage{siunitx}
\usepackage[numbers, square]{natbib}
\usepackage{pgfplots}
\usepackage{subcaption}
\pgfplotsset{compat=newest}
\usepackage{multirow}
\usepackage{authblk}

\graphicspath{{./figs/}}

\title{Multigrid on unstructured meshes with regions of low quality cells}

\author[1]{Yuxuan Chen\thanks{\url{chenyx@suda.edu.cn},
\url{https://orcid.org/0000-0002-4152-2011}}}
\author[2]{Garth N.~Wells\thanks{\url{gnw20@cam.ac.uk},
\url{https://orcid.org/0000-0001-5291-7951}}}

\affil[1]{School of Mathematical Sciences, Soochow University, Suzhou, China}
\affil[2]{Department of Engineering, University of Cambridge, Cambridge, United Kingdom}

\date{}
\begin{document}
\maketitle
\begin{abstract}
\noindent
The convergence of multigrid methods degrades significantly if a small
number of low quality cells are present in a finite element mesh, and
this can be a barrier to the efficient and robust application of
multigrid on complicated geometric domains. The degraded performance is
observed also if intermediate levels in a non-nested geometric multigrid
problem have low quality cells, even when the fine grid is high quality.
It is demonstrated for geometric multigrid methods that the poor
convergence is due to the local failure of smoothers to eliminate parts
of error around cells of low quality. To overcome this, a global--local
combined smoother is developed to maintain effective relaxation in the
presence of a small number of poor quality cells. The smoother involves
the application of a standard smoother on the whole domain, followed by
local corrections for small subdomains with low quality cells. Two- and
three-dimensional numerical experiments demonstrate that the degraded
convergence of multigrid for low quality meshes can be restored to the
high quality mesh reference case using the proposed smoother. The effect
is particularly pronounced for higher-order finite elements. The results
provide a basis for developing efficient, non-nested geometric multigrid
methods for complicated engineering geometries.
\\*[2ex]
\end{abstract}

\section{Introduction}
\label{sec:introduction}

Multigrid methods have the potential to be optimal solvers ($O(n)$
complexity, where $n$ is the number of unknowns) for systems arising in
the solution of elliptic partial differential
equations~\cite{briggs2000multigrid,hackbusch2013multi,trottenberg2000multigrid},
and lend themselves to efficient parallel implementations. This makes
tractable high fidelity simulations of complex engineering components
and enables the computation of engineering problems at a system level.

Finite element simulations on complex geometries are invariably
performed on unstructured grids, and cell quality will vary with
position. This is particularly the case when representing geometrically
complex shapes with modest cell counts. It is, however, well recognised
that cell quality can have a significant impact on the performance of
iterative solvers \cite{shewchuk2002good,freitag2000cost,katz2011mesh}.
When disappointing performance of multigrid solvers is observed in
engineering practice (slow performance with high iteration counts, or
failure to converge), low cell quality is a common
cause~\cite{richardson:2018}. Even very small regions with just a few
low quality cells can cause convergence of a multigrid preconditioned
iterative solver to stall. On the other hand, it has been shown that if
using linear Lagrange finite elements for solving Poisson equation, the
standard \emph{a priori} estimate may remain valid on a mesh with low
quality regions \cite{duprez2018finite}, but the associated finite
element linear system is poorly conditioned.

Ideally, computational grids would be of sufficient quality to not
degrade performance of the linear solver. However, we consider two
contexts in which this is not universally feasible. The first is that
the generation of meshes for highly complex geometries and for which
ensuring that all cells are of high quality may be very difficult.
Considerable research has been devoted to improving mesh quality,
e.g.~\cite{knupp2000achieving1,knupp2000achieving2,klingner2008aggressive}.
However, even in cases where it is technically possible to create a high
quality mesh, from a workflow perspective it may be more efficient (less
costly in overall time) if the solver can deal sub-optimal grids
efficiently. The second case is geometric (non-nested) multigrid, in
which the quality of intermediate coarse grids may be compromised by
under-resolution with respect to the domain shape complexity, i.e.~a
low-resolution triangulation of a complex domain will necessarily
compromise on cell quality. Ideally the fine grid will be of high
quality to provide good approximation properties, but the coarse grids
are primarily a vehicle for constructing a (hopefully) fast solver.
Robust performance of a geometric multigrid solver with respect to the
quality of intermediate grids is appealing for complex engineering
geometries.

We examine the performance of a geometric multigrid method for the
finite element method on non-nested unstructured meshes in the presence
of a small number of low quality cells. It is shown, through examples,
that standard multigrid smoothers fail to eliminate parts of the error
in highly localised regions around small clusters of low quality cells.
The component of the residual that is not reduced by the smoother can be
`lost' in the restriction operation. Building on the observation of how
smoothers perform around low quality cells, we construct a global--local
combined smoother in which (i) a standard multigrid smoother is applied
over the entire grid, followed by (ii) a local correction on small
regions with low cell quality using a direct solver. The combined
smoother is effectively a Schwarz-type domain decomposition method with
full overlap \cite{smith2004domain,dolean2015introduction}. We assume
that the number of unknowns in each low quality region is small such
that the cost of applying a direct solver on these regions is small
relative to other operations in the solver. In this work, we assume
conforming boundaries between geometric multigrid levels. Examples of
handling non-conforming boundaries in geometric multigrid
include~\cite{mavriplis1992three,smith1993multigrid,dickopf2010multilevel}.

The remainder of this paper is structured as follows. In
\cref{sec:background}, we give a brief overview of multigrid. An
explanation of why the performance of multigrid degrades in the presence
of low quality cells is presented in \cref{sec:reason}, supported by
numerical examples. A global--local combined smoother is introduced in
\cref{sec:local}, followed by a brief detour in \cref{sec:chebyshev} to
discuss a specific issue for Chebyshev smoothers. Numerical examples are
presented in \cref{sec:example} for Poisson and elasticity problems.
Conclusions are drawn in \cref{sec:conclusion}.

\section{Multigrid background}
\label{sec:background}

Consider a domain $\Omega \subset \mathbb{R}^{\dim}$, where $\dim = 1,
2, 3$, and triangulations (grids) of the domain $\Omega_l$, $l = 1, 2,
\ldots, L$. With increasing index $l$ the grids become coarser, i.e.~the
number of cells reduces. We wish to solve a finite element problem on
the finest grid $\Omega_{1}$. The finite element method generates the
discrete operator $A_{1} \in \mathbb{R}^{n \times n}$, and the task is
to solve the linear system
\begin{equation}
  A_{1} u_{1} = b_{1},
\label{eqn:system_finest}
\end{equation}
where $b_1\in \mathbb{R}^n$ is the right-hand side vector and $u_{1} \in
\mathbb{R}^n$ is the vector of degrees-of-freedom.

If $V_{l}$ is a finite element space on $\Omega_{l}$, in the case of
nested grids and using the same element type on each level we have
$V_{l+1} \subset V_l$. With a view to complex geometries, we consider in
this work non-nested meshes, i.e.~$V_{l+1} \not \subset V_l$. A
\emph{prolongation} operator $P_l: V_{l + 1} \rightarrow V_l$ projects a
finite element function on grid $\Omega_{l + 1}$ onto the next finest
grid $\Omega_{l}$. We consider continuous Lagrange finite element spaces
and define the prolongation operator using interpolation, in which case
the components of $P_l$ are given by
\begin{equation}
  [P_l]_{ij} = \varphi^{(l+1)}_{j}(x_i),
\end{equation}
where $\varphi^{(l+1)}_{j}$ is the basis function associated with the
$j$th degree-of-freedom on level $l+1$ and $x_i$ is the interpolation
coordinate for the $i$th basis function on the finer $\Omega_{l}$ grid.
The \emph{restriction} operator $R_{l}$ maps a function on grid $l$ to
the coarser $l + 1$ grid. Following the Galerkin approach, we set $R_{l}
= P_{l}^{T}$, leading to
\begin{equation}
  A_{l+1} := P_{l}^{T} A_{l} P_{l}.
\end{equation}
for the operator on level $l + 1$, $A_{l+1}$, and
\begin{equation}
  b_{l+1} := P_{l}^{T} \left(b_l-A_lu_l \right)
\end{equation}
for the residual vector on level $l + 1$, $b_{l+1}$.

A \emph{smoother} $S$ provides an approximate solution $u_{l}$ to the
problem $A_{l} u_{l} = b_{l}$:
\begin{equation}
  u_{l} \leftarrow S^{\nu}(A_l,b_l,u_l),
\end{equation}
where $\nu$ denotes the number of applications of the smoother. Commonly
used smoothers include (weighted) Jacobi, (symmetric) Gauss--Seidel and
Chebyshev iterations. For the coarsest grid, $\Omega_{L}$, a direct
solver is employed. Given the fine grid operator $A_{1}$, the fine grid
right-hand side vector $b_{1}$ and the prolongation operators, a V-cycle
Galerkin multigrid algorithm with $L$ levels for solving
\cref{eqn:system_finest} is summarised in \cref{alg:mg}.
\begin{algorithm}
\caption{Multigrid V-cycle of $L$ levels to solve $A_1u_1=b_1$.}
\label{alg:mg}
\begin{algorithmic}[1]
\Procedure{$u_{l}\leftarrow \textrm{Vcycle}(A_l,b_l,u_l, l,\nu)$}{}
\For {$l=1,2,\cdots,L$}
\If {$l\neq L$}
\State Pre-smoothing $u_{l} \leftarrow S^{\nu}(A_l,b_l,u_l)$.
\State Coarse grid construction $A_{l+1}=P_l^T A_l P_l$, $b_{l+1}=P_l^T (b_l-A_lu_{l})$ and $u_{l+1}=P_l^T u_l$.
\Else
\State Direct solver on the coarsest grid for $A_{L} u_{L} = b_{L}$.
\EndIf
\EndFor
\For {$l=L-1,L-2,\cdots,1$}
\State Updating current solution $u_l\leftarrow u_l+P_lu_{l+1}$.
\State Post-smoothing $u_{l}\leftarrow S^{\nu}(A_l,b_l,u_l)$.
\EndFor
\EndProcedure
\end{algorithmic}
\end{algorithm}

We consider a hierarchy of non-nested grids, but restricted to the case
where all grids conform to the same boundary. A natural extension would
be to non-conforming boundaries, which would require some additional
considerations
\cite{mavriplis1992three,smith1993multigrid,dickopf2010multilevel}.

\section{Why multigrid converges slowly with low quality meshes}
\label{sec:reason}

We explore how multigrid behaves with non-nested unstructured grids when
levels have small regions of low quality cells. There is no universal
measure of cell quality, and no sharp distinction between `low' and
`high' quality cells. Some cell quality measures are discussed in
\cite{parthasarathy1994comparison, knupp2001algebraic}. We consider
simplex cells and select the radius ratio as a measure of cell quality;
the radius ratio is the ratio of a cell's inscribed sphere radius
($R_I$) to its circumscribing sphere radius ($R_C$). We define a
normalised radius ratio $\gamma$ as:
\begin{equation}
  \gamma := \frac{R_I}{\gamma^{*} R_C},
  \label{eqn:radius_ratio}
\end{equation}
where $\gamma^*$ is the optimal radius ratio, which is equal to the
geometric dimension. The normalised radius ratio is in the range $(0,
1]$, with ideal cells having a measure of $1$, and degenerate cells
having the measure~$0$.

\subsection{Model problem}
\label{sec:model}

We solve the homogeneous Poisson problem on the unit square $\Omega
=[0,1]^{2}$,
\begin{equation}
  \begin{aligned}
  -\nabla^2 u&=0 \quad \text{in} \ \Omega,
  \\
  u&=0 \quad \text{on} \ \partial \Omega,
  \end{aligned}
\label{eqn:poisson}
\end{equation}
using the finite element method with degree one Lagrange basis functions
and a two-level multigrid solver (V-cycle) with unstructured, non-nested
meshes. The coarse grid contains $68$ cells and the fine grid contains
$272$ cells. The initial solution guess interpolates $u^{(0)} = \sin(10
\pi x) \sin(10 \pi y)$.

\begin{figure}
  \centering
  \includegraphics[width=0.4\linewidth]{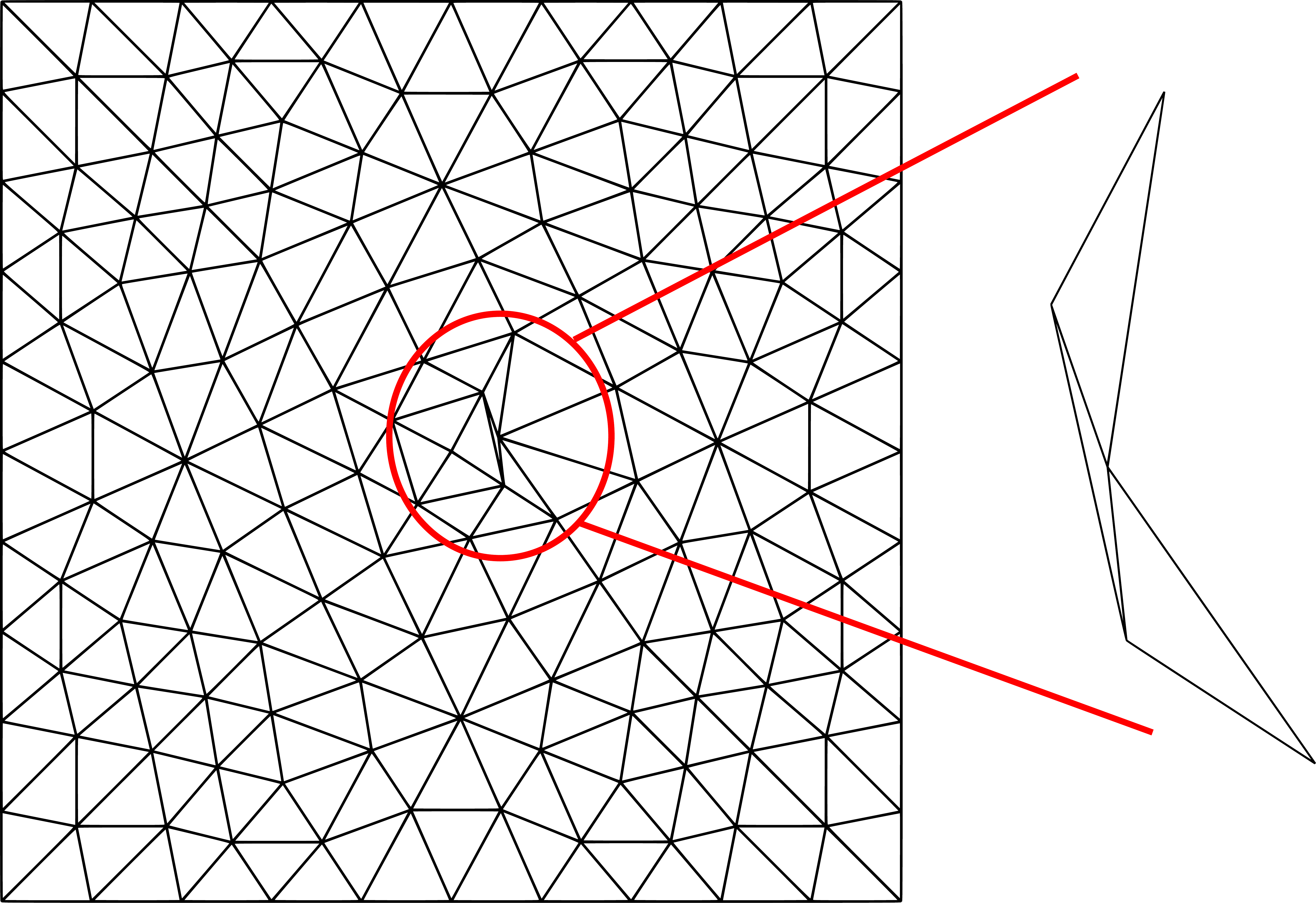}
  \caption{Unstructured grid of the unit square with a poor quality
  region near the centre.}
  \label{fig:squaremesh1}
\end{figure}

\begin{figure}
  \centering
  \includegraphics[width=0.4\linewidth]{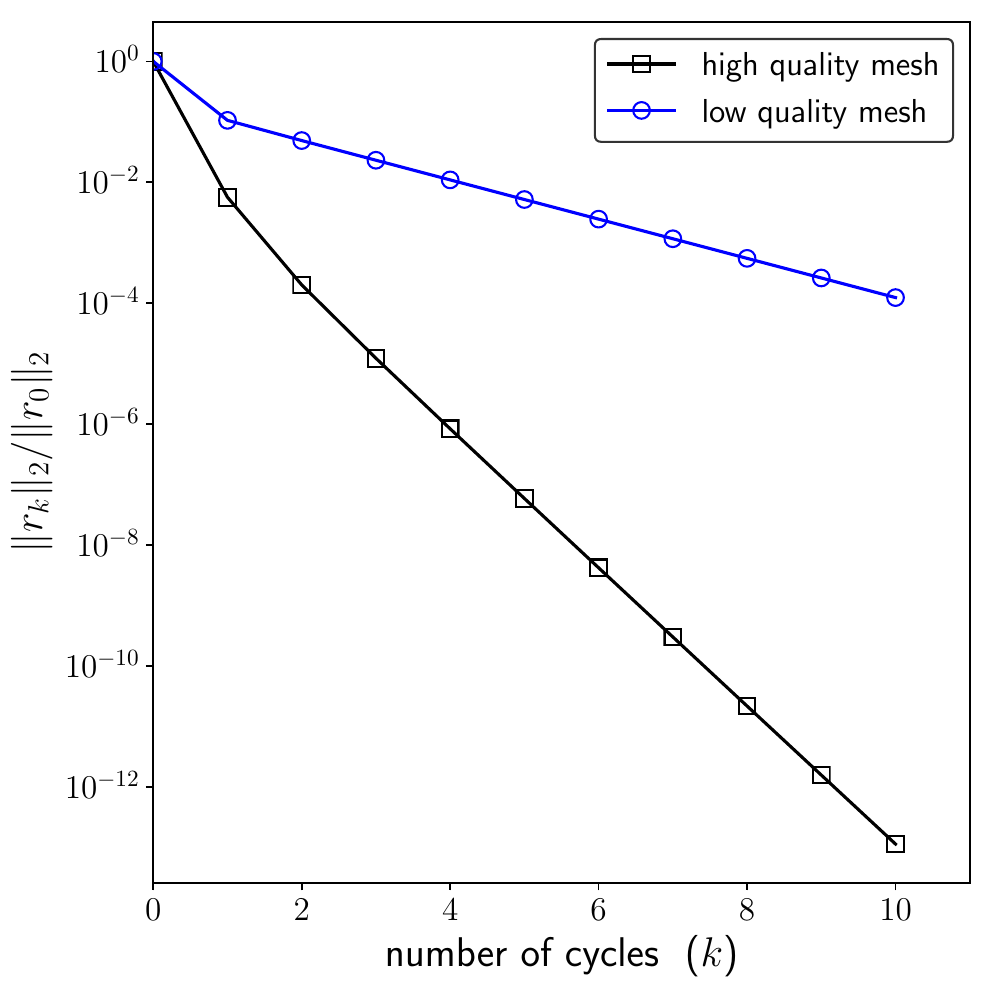}
  \caption{Relative residual at the end of each multigrid cycle for high
  and low quality meshes. Symmetric Gauss--Seidel is used as the
  smoother.}
  \label{fig:square1_bad}
\end{figure}

Two cases are considered: the first uses a high quality fine grid, and
the second uses a perturbation of the high quality fine grid to create a
small region of low quality cells. The perturbed fine grid is shown in
\cref{fig:squaremesh1}. The minimum angle over all cells in the
perturbed fine grid is approximately $\pi / 180$, corresponding to a
normalised radius ratio of $10^{-3}$ for the cell. There are three `low'
quality cells with a normalised radius ratio of less than~$0.1$.
For a symmetric Gauss--Seidel smoother (one iteration at each
application of the smoother, pre- and post-smoothing), the relative
residual after each cycle is recorded and presented in
\cref{fig:square1_bad}. The convergence rate for the low quality fine
grid case is dramatically slower than the high quality case.

\subsection{Local failure of smoothers}
\label{sec:smoother}

\begin{figure}
  \centering
  \begin{subfigure}[t]{.3\textwidth}
    \centering
    \includegraphics[width=0.8\linewidth]{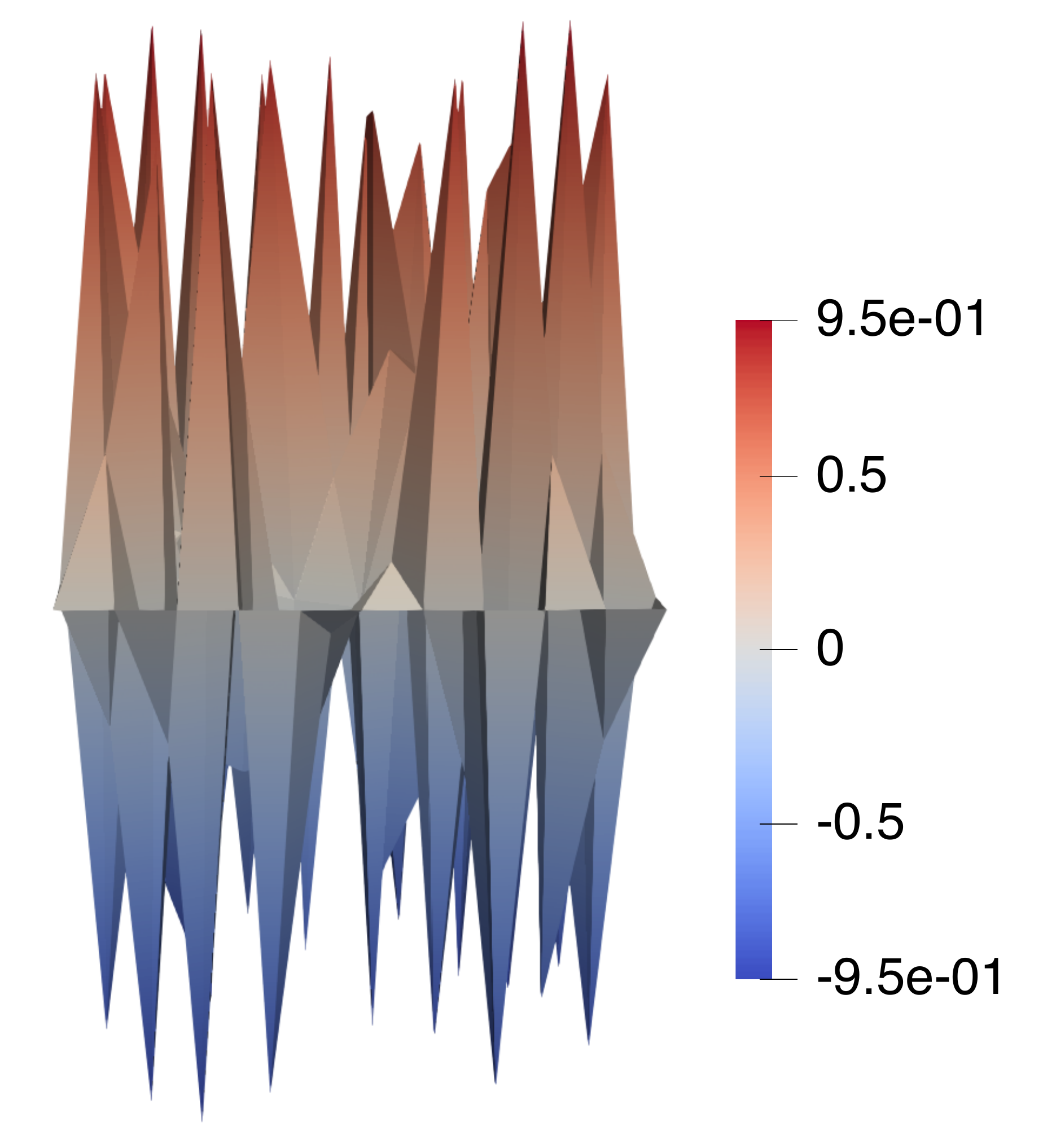}
    \caption{Initial guess $u^{(0)}$.}
    \label{fig:initial}
  \end{subfigure}%
  \begin{subfigure}[t]{.5\textwidth}
    \centering
    \includegraphics[width=0.8\linewidth]{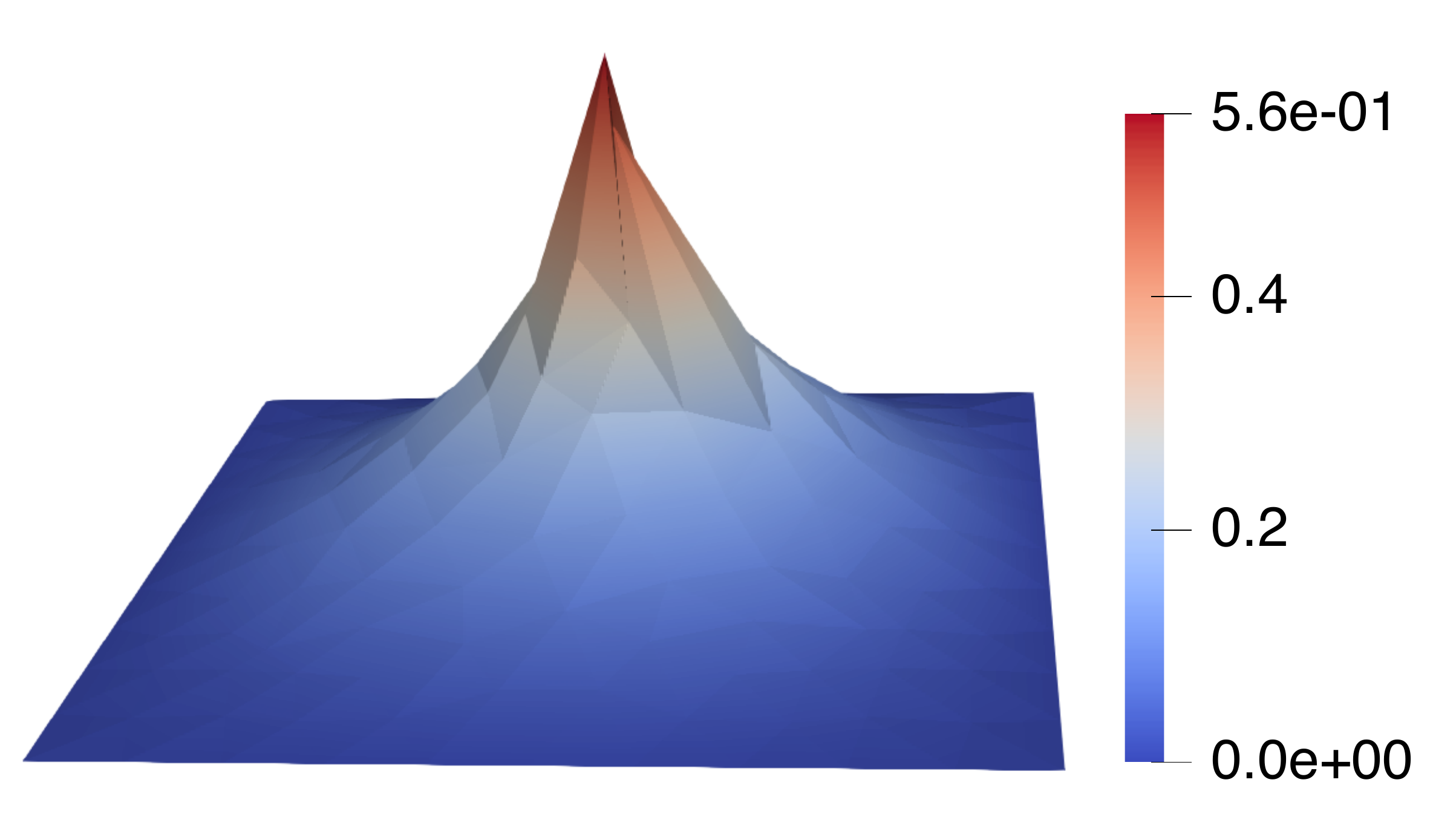}
    \caption{Absolute value of error on each vertex.}
    \label{fig:gauss_five}
  \end{subfigure}
  \caption{A (a) high frequency Fourier mode as initial guess, and (b)
  absolute value of the error at each vertex after five iterations of
  symmetric Gauss--Seidel.}
  \label{fig:gauss_error}
\end{figure}

We investigate the performance of a smoother only for the model problem
on the grid shown in \cref{fig:squaremesh1}. The interpolation of the
initial guess $u^{(0)}$ is shown in \cref{fig:initial}, and the absolute
value of the error after five symmetric Gauss--Seidel iterations is
shown in \cref{fig:gauss_five}. After applications of the smoother, in
most of the domain the oscillatory error has been eliminated. However, a
localised error persists around the low quality cells.

\begin{figure}
  \centering
  \includegraphics[width=0.4\linewidth]{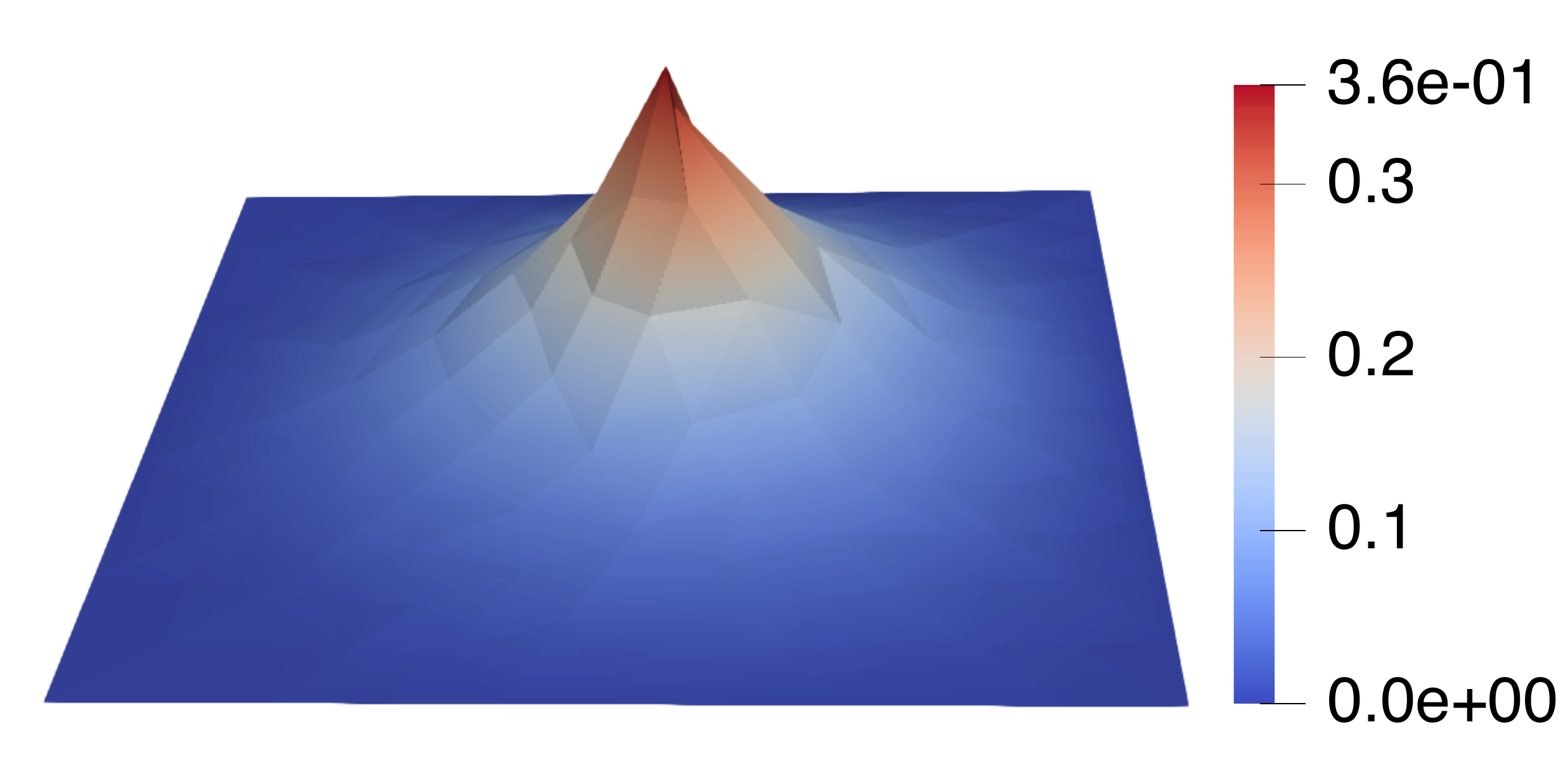}
  \caption{Absolute value of error on each vertex after five iterations
  of Jacobi-preconditioned Chebyshev method on the low quality unit
  square model problem.}
\label{fig:chebyshev_error}
\end{figure}

We also apply a Jacobi--preconditioned Chebyshev smoother
\cite{adams2003parallel} to the model problem. The largest eigenvalue
used in the Chebyshev method is estimated by the Krylov--Schur method
\cite{stewart1,hernandez2007krylov} (tolerance set as $10^{-8}$), and
the smallest eigenvalue is set as one tenth of the largest one. The
absolute value of the error at each vertex after five applications of
the smoother is shown in \cref{fig:chebyshev_error}. As with the
Gauss--Seidel smoother, the localised error persists in the region of
low cell quality.

The observation from the model problem is that for a mesh with a small
region of low quality cells, standard multigrid smoothers are
ineffective in the region around the low quality cells. The error is not
eliminated and may subsequently be lost in the restriction operator to
the coarser grid, and this slows the rate of convergence of multigrid.
The localised nature of the residual error motivates the consideration
of local smoothers/solvers that can provide effective smoothing in small
regions.

\section{Global--local correction smoother}
\label{sec:local}

We consider a global--local combined smoother that overcomes the issues
demonstrated in the previous section. We consider the linear system $Au
= b$, where $A\in \mathbb{R}^{n \times n}$. Let $u^{(k)}$ be the
approximate solution obtained after $k$ iterations of some iterative
method, and $r^{(k)} := b - A u^{(k)}$ be the corresponding residual.

\subsection{Local residual correction}

Suppose there are $D$ small, closed subdomains containing low quality
cells, which we denote by $\Omega_B^d \subset \Omega, \ d = 1, 2,
\ldots, D$. The subdomains are defined to be disjoint, $\bigcap_{d}
\Bar{\Omega}_B^d = \emptyset$. Let $B_d$ be the set of all
degrees-of-freedom (DOFs) $\beta_i$ associated with the closure of the
subdomain $\Bar{\Omega}_{B}^d$, i.e., $B_d = \{\beta_i : i = 1, 2,
\ldots, n_B^d\}$, where $n_B^d := \lvert B_d \rvert$. Let $I_d \in
\mathbb{R}^{n \times n_B^d}$ be the natural inclusion mapping
degrees-of-freedom in $\Bar{\Omega}_B^d$ to the the whole domain
$\Omega$, given by:
\begin{equation}
  [I_d]_{ij}
  :=
  \begin{cases}
    1 \quad \quad i=\beta_j, \, \beta_j \in B_d,
    \\
    0 \quad \quad \text{otherwise}.
  \end{cases}
\end{equation}

Consider a restriction of the matrix $A$ to subdomain
$\Bar{\Omega}_{B}^{d}$,
\begin{equation}
  A^{[d]} := I_{d}^{T} A I_{d},
  \label{eqn:local_system}
\end{equation}
which has size $n_B^d \times n_B^d$. Similarly for residual,
\begin{equation}
  r^{[d]} := I_d^T r^{(k)}.
  \label{eqn:local_residual}
\end{equation}
A local residual correction system on subdomain~$\Bar{\Omega}_B^d$ is then
given by:
\begin{equation}
  A^{[d]} e^{[d]} = r^{[d]},
\label{eqn:local}
\end{equation}
where $e^{[d]}$ is the local error correction. If the original system
$A$ is symmetric positive-definite, then the local system $A^{[d]}$ is
symmetric positive-definite \citep[Lemma~3.1]{xu1992iterative}. We
assume that the local correction systems are small and can be solved
efficiently using a direct solver:
\begin{equation}
  e^{[d]} = (I_d^T A I_d)^{-1} I_d^T r^{(k)}.
\end{equation}
The local error correction can be mapped back to the global domain by
applying~$I_{d}$ and the approximate solution $u^{(k)}$ corrected,
\begin{equation}
  u^{(k+1)} = u^{(k)} + \sum\limits_{d=1}^{D} I_d e^{[d]}.
  \label{eqn:local_update}
\end{equation}

The local residual correction can be written in a preconditioner form as
\begin{equation}
  u^{(k+1)} = u^{(k)} + S_c \left(b - Au^{(k)}\right),
\end{equation}
where
\begin{equation}
  S_c := \sum\limits_{d=1}^{D}I_d(I_d^T A I_d)^{-1} I_d^T.
\end{equation}
In the case of an interior region of low quality cells, the local
correction is essentially the solution of a local Dirichlet problem.

The local residual correction procedure is summarised in
\cref{alg:local}.
\begin{algorithm}
\caption{Local residual correction $S_c$}
\label{alg:local}
\begin{algorithmic}
\Procedure{$u^{(k+1)}=S_{c}(A,b,u^{(k)})$}{}
\State Identify low quality regions $\Omega_B^d$
and the corresponding DOF sets $B_d$, $d=1,2,\cdots, D$.
\For{$d=1,2,\cdots,D$}
\State Construct the local residual correction systems $A^{[d]}$, $r^{[d]}$
via \cref{eqn:local_system} and \cref{eqn:local_residual}.
\State Solve the local residual correction system $A^{[d]} e^{[d]} = r^{[d]}$ by a direct method.
\State Correct $u^{(k)}$ by adding local errors via \cref{eqn:local_update}.
\EndFor
\EndProcedure
\end{algorithmic}
\end{algorithm}
It is straightforward to implement and the additional computational cost
is small if the poor quality regions are few and small. It is noted that
the local residual correction is similar to subspace correction methods,
including the parallel subspace correction and the successive subspace
correction, proposed by Xu \cite{xu2001method,xu1992iterative}. In
particular, the low quality regions can be viewed as (local) subdomains.
The local correction $S_c$ corrects the error on the local subdomains,
and can be interpreted as a parallel subspace correction on the
subdomains with poor quality cells.

\subsection{A global--local combined smoother for multigrid}

We consider a global--local combined multigrid smoother, with the local
smoother applied in a number of small regions to overcome the
deleterious effect of any low quality cells. The smoother has a
`sandwich' form, with the local correction smoother $S_{c}$ applied on
subdomains $\Omega_B^d$, followed by a standard global smoother $S_g$,
e.g.~a symmetric Gauss--Seidel, on the whole domain/level, followed by
another application of the local smoother $S_{c}$. The smoother involves
three steps:
\begin{equation}
  \begin{aligned}
  u^{(k+1/3)} &= u^{(k)}+S_c \left( b-Au^{(k)} \right),
  \\
  u^{(k+2/3)} &= u^{(k+1/3)}+S_g \left( b-Au^{(k+1/3)} \right),
  \\
  u^{(k+1)} &= u^{(k+2/3)}+S_c \left( b-Au^{(k+2/3)} \right).
  \end{aligned}
\end{equation}
Rearranging,
\begin{equation}
  u^{(k+1)} = u^{(k)} + S_{gc} \left(b - Au^{(k)}\right),
\end{equation}
where
\begin{equation}
  S_{gc} = 2S_c - S_cAS_c + (I - S_c A) S_g (I-AS_c).
\end{equation}
demonstrating that the smoother is symmetric.

\subsection{Relationship to a domain decomposition method}
\label{sec:relation}

\begin{figure}
  \centering\includegraphics[width=0.7\textwidth]{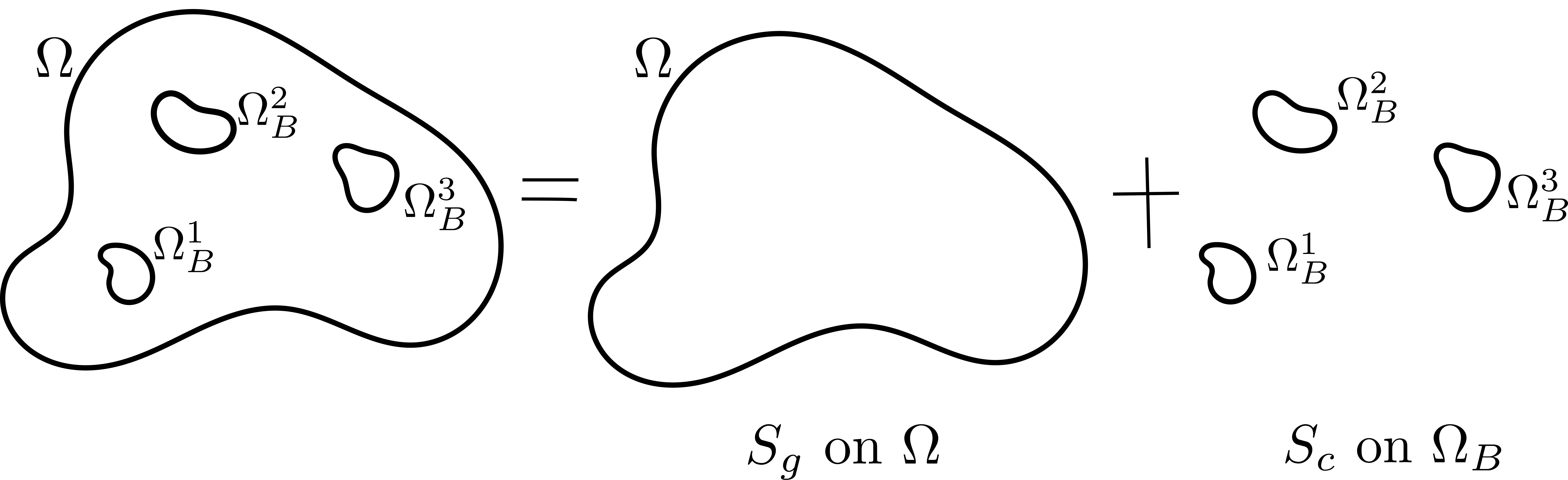}
  \caption{Domain decomposition view of the combined global--local
  smoother, which consists of the global smoother $S_g$ on the whole
  domain and the local residual correction $S_c$ on the locally poor
  quality subdomains.}
  \label{fig:decomposition}
\end{figure}

The global--local combined smoother can be viewed as a domain
decomposition by considering two domains: the whole domain and several
small subdomains, as shown in \cref{fig:decomposition}. Taking the whole
domain as the first `subdomain', then the domain decomposition is based
on
\begin{equation}
  \Omega=\Omega \, \cup \, \{\Omega_B^1 \, \cup \, \cdots \,\cup \, \Omega_B^D\}
    =\Omega \, \cup \Omega_B.
\end{equation}
\Cref{fig:decomposition} illustrates the domain decomposition view of
the global--local smoother.

This combined smoother can be viewed as a full overlapped Schwarz-type
domain decomposition method
\cite{smith2004domain,dolean2015introduction}. It is also noted in
\cite{dolean2015introduction} that for the Schwarz-type domain
decomposition method the larger the subdomain overlap the faster the
convergence. From the algebraic point of view, the combined smoother is
also in the form of a block Gauss--Seidel. The combined smoother is
similar to a plane smoother \cite{trottenberg2000multigrid,
llorente2000behavior, llorente2000alternating}, which is a variant of
block Gauss--Seidel. The plane smoother applies a standard smoother on
multi-block structured grids, e.g.~on $x$- or $y$-planes and is known
for its effectiveness for strongly anisotropic problems
\cite{oosterlee1997gmres}. It is also noted that in
\cite{Farrell1,Farrell2} a multigrid preconditioner is proposed that
solves systems on local patches for the augmented momentum block of a
finite element discretisation of the incompressible Navier--Stokes
equations.

\subsection{Identifying low quality regions}
\label{sec:identify}

There is no universal or binary measure of cell quality, so we
heuristically select a cell quality measure and set a threshold value
for what constitutes a low quality cell. For the examples in
\cref{sec:example}, which use simplex cells, we define low quality
regions $\Omega_B$ by:
\begin{equation}
\begin{aligned}
  \Omega_b &= \{K: \ \gamma(K)< 0.1\},\\
  \Omega_B &=\Omega_b \cup \{K: K \text{ shares vertex with } \Omega_b\},
\label{eqn:identify}
\end{aligned}
\end{equation}
where the region $\Omega_b$ contains all cells $K$ with a normalised
radius ratio $\gamma$ less than a specified threshold (see
\cref{eqn:radius_ratio} for the definition of the radius ratio). In the
case of Lagrange elements, the entire local correction region $\Omega_B$
includes the region $\Omega_b$ and its one layer (by vertex) extension,
which involves all cells sharing degrees-of-freedom with~$\Omega_b$.

\begin{figure}
  \centering
  \includegraphics[width=0.7\textwidth]{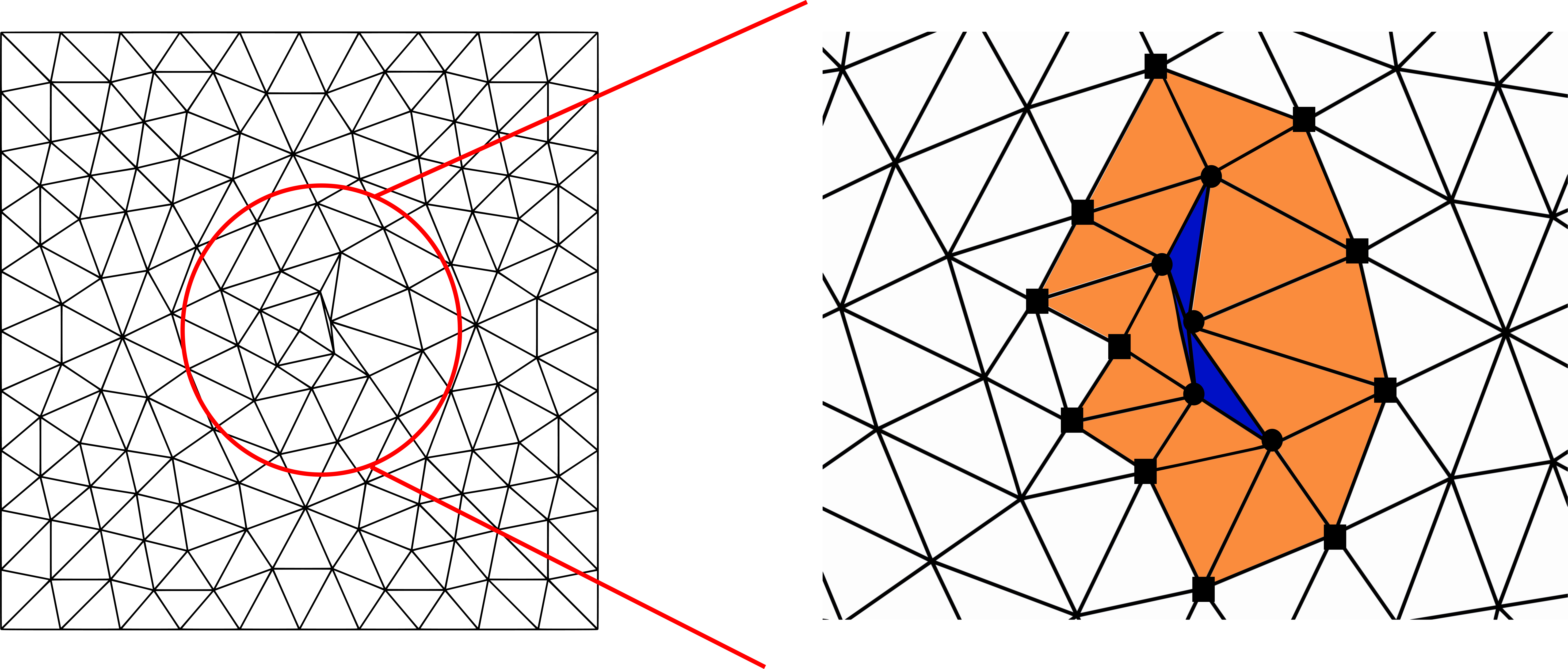}
  \caption{Local correction region $\Omega_B$ in the low quality unit
  square mesh (\cref{fig:squaremesh1}). The area coloured blue is
  $\Omega_b$, the one-layer extension of $\Omega_b$ is coloured orange.
  The local correction region $\Omega_B$ is the union of coloured
  cells.}
  \label{fig:identify}
\end{figure}

\Cref{fig:identify} illustrates the local correction region for the mesh
in \cref{fig:squaremesh1}, with low quality cells being those with
normalised radius ratio $\gamma < 0.1$ in this case. The low quality
cells in $\Omega_b$ are coloured blue. The cells in the one layer
extension are coloured orange. The local correction region
$\Omega_B$ is the entire coloured region.

\section{Chebyshev smoothers: adjusted eigenvalues}
\label{sec:chebyshev}

The Chebyshev semi-iterative method \cite{hageman2012applied} is
commonly used as a multigrid smoother \cite{adams2003parallel}, and
requires knowledge of the largest and smallest eigenvalues of the
operator. For effective application as a smoother, only the largest
eigenvalue needs to be approximated \cite{adams2003parallel}. In
practice, when used as smoother the estimated smallest eigenvalue is set
to be specified fraction of the estimated largest eigenvalue.

In the presence of localised low quality cells, the maximum eigenvalue
increases relative to a comparable resolution high quality mesh. For a
mesh with a small region of low quality cells, experiments indicate that
the increased maximum eigenvalue may render the smoother less effective
in the bulk of the domain. Consider a block-wise decomposition of the
discretised system
\begin{equation}
  A
  =
  \begin{bmatrix}
    A_{gg} & A_{gb} \\
    A_{bg} & A_{bb}
  \end{bmatrix},
\end{equation}
where $A_{bb}$ corresponds to the DOFs associated with the closure of
$\Omega_B$. The \emph{adjusted eigenvalue} is taken as the largest
eigenvalue of $A_{gg}$, i.e., $\lambda_{\max} \leftarrow
\lambda_{\max}(A_{gg})$. We explore numerically in \cref{sec:example}
differences in performance using estimates of the largest eigenvalue of
$A$ compared to the adjusted largest eigenvalue.

\section{Numerical examples}
\label{sec:example}

We examine numerically the performance of non-nested geometric Galerkin
multigrid with the proposed smoother for model Poisson and elasticity
problems. Problems are solved using Lagrange elements on simplices using
linear ($P_1$) and quadratic ($P_2$) bases. The solver is terminated
once the relative residual reaches $10^{-10}$, measured in the
$\ell^{2}$-norm. For Chebyshev smoothers, the Krylov--Schur method
\cite{stewart1, hernandez2007krylov} is used to compute the largest
eigenvalue, using a tolerance of~$10^{-8}$.

All meshes (grids) are unstructured and the levels are non-nested. We
restrict ourselves to problems in which the geometry can be exactly
represented by the coarsest grid. To study the influence of cell
quality, we displace some vertices in the generated meshes to change,
controllably, mesh quality characteristics. Low quality regions
$\Omega_B$ are identified as regions with normalised cell radius ratios
of less than $0.1$ and their one layer extension (given in
\cref{eqn:identify}). We use $\gamma_{\min}$ to denote the minimum
normalised radius ratio in a mesh (grid). When distinguishing between
high- and low-quality meshes we use the annotation
$\Omega_{\text{high}}$ and $\Omega_{\text{low}}$, respectively.

We consider two cases:
\begin{trivlist}
\item {\bf Case A}: All grids (meshes) have low quality regions.
\item {\bf Case B}: All grids below the finest grid have low quality
regions. The finest grid is high quality.
\end{trivlist}
Reference performance is taken as the performance for problems with high
quality meshes for all levels.

The examples are computed using libraries from the FEniCS Project
\cite{alnaes2015fenics,logg2010dolfin,logg2012automated}
and PETSc \cite{petsc-web-page,petsc-user-ref,petsc-efficient}. The
Galerkin finite element systems and the restriction/prolongation
operators are constructed with FEniCS, and the remaining generic
multigrid functionality is provided by PETSc. The largest eigenvalue,
used in the Chebyshev smoother, is estimated using the SLEPc library
\cite{Hernandez:2005:SSF,slepc-users-manual}. Meshes for each level are
generated using Gmsh \cite{geuzaine2009gmsh}. The source code is freely
available in the supporting material \cite{supporting-material}.

\subsection{Poisson problem}

For a domain $\Omega \subset \mathbb{R}^{d}$, $d = 2, 3$ with boundary
$\Gamma := \partial\Omega$ that is partitioned such that $\Gamma =
\Gamma_{D} \cup \Gamma_{N}$ and $\Gamma_{D} \cap \Gamma_{N} =
\emptyset$, the Poisson problem reads:
\begin{equation}
  \begin{aligned}
    - \nabla^2 u &= f\quad \text{in} \ \Omega,
  \\
  u &= g \quad \text{on} \ \Gamma_D,
  \\
  \nabla u \cdot n &= s\quad \text{on} \ \Gamma_N.
  \end{aligned}
\end{equation}
where $f$, $g$ and $s$ are prescribed. We examine performance for two-
and three-dimensional problems. Multigrid is used as a solver for the
Poisson examples (as opposed to as a preconditioner).

\subsubsection{Unit square domain}

\begin{figure}
  \centering
  \includegraphics[width=0.5\textwidth]{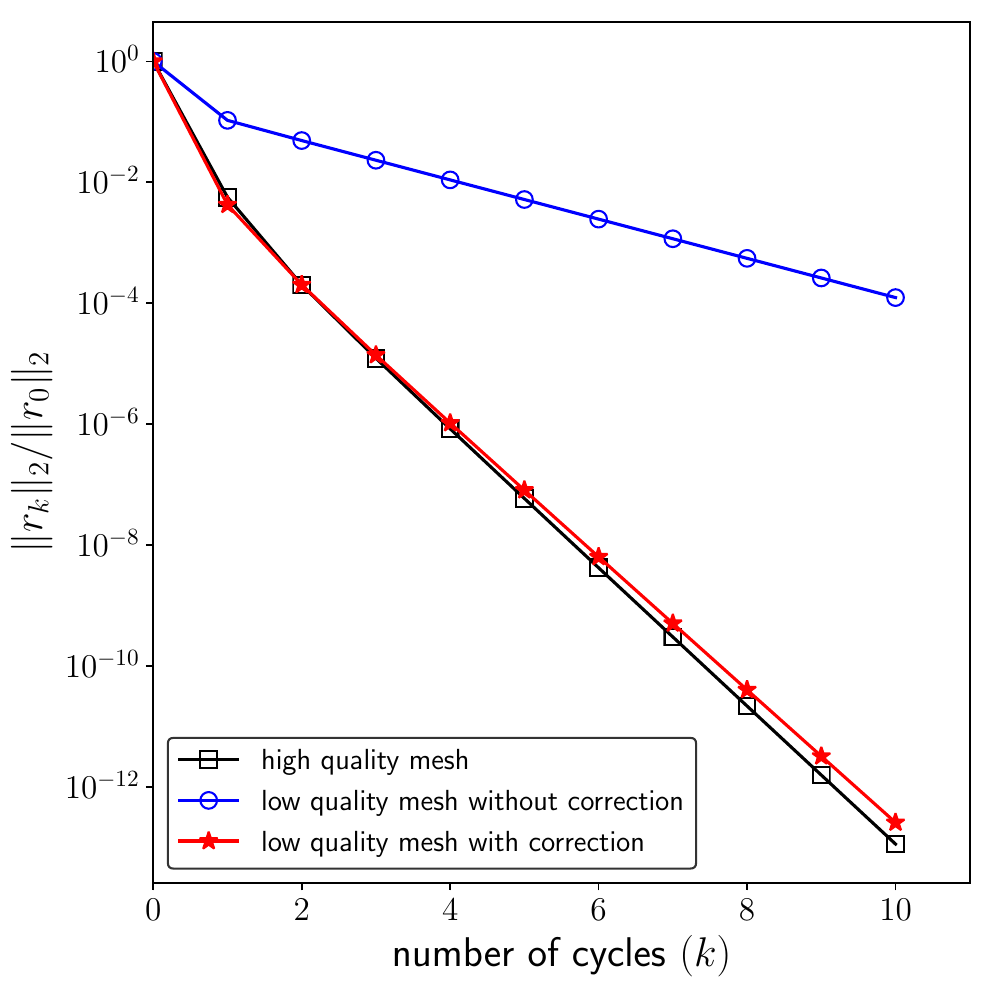}
  \caption{Relative residual with and without the local correction for
  the Poisson problem on a unit square.}
\label{fig:res_square1}
\end{figure}

\begin{figure}
  \centering
  \begin{subfigure}[t]{0.4\textwidth}
    \centering
    \includegraphics[width=0.8\linewidth]{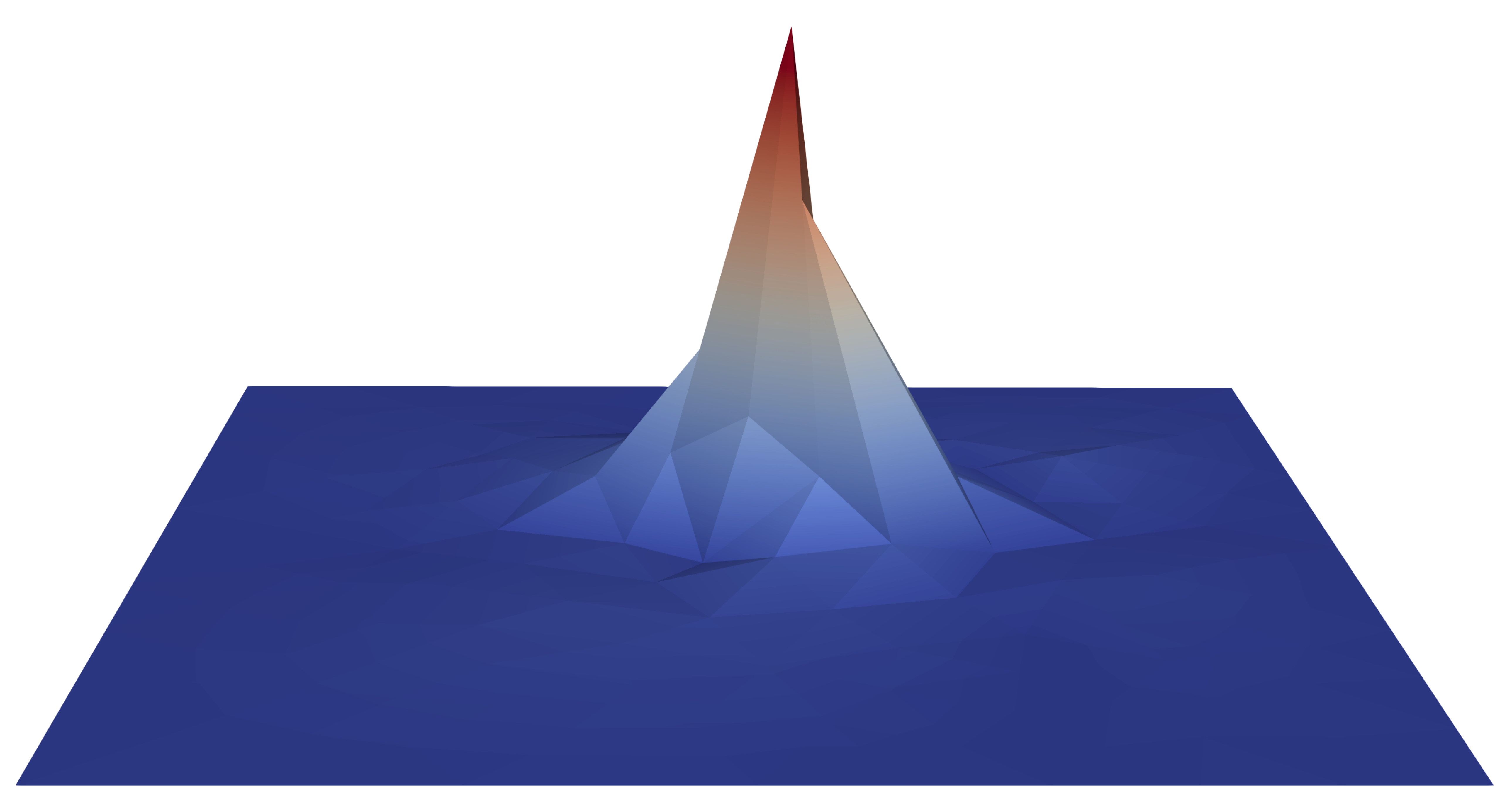}
    \caption{without local correction}
  \end{subfigure}\hfill%
  \begin{subfigure}[t]{0.4\textwidth}
    \centering
    \includegraphics[width=0.8\linewidth]{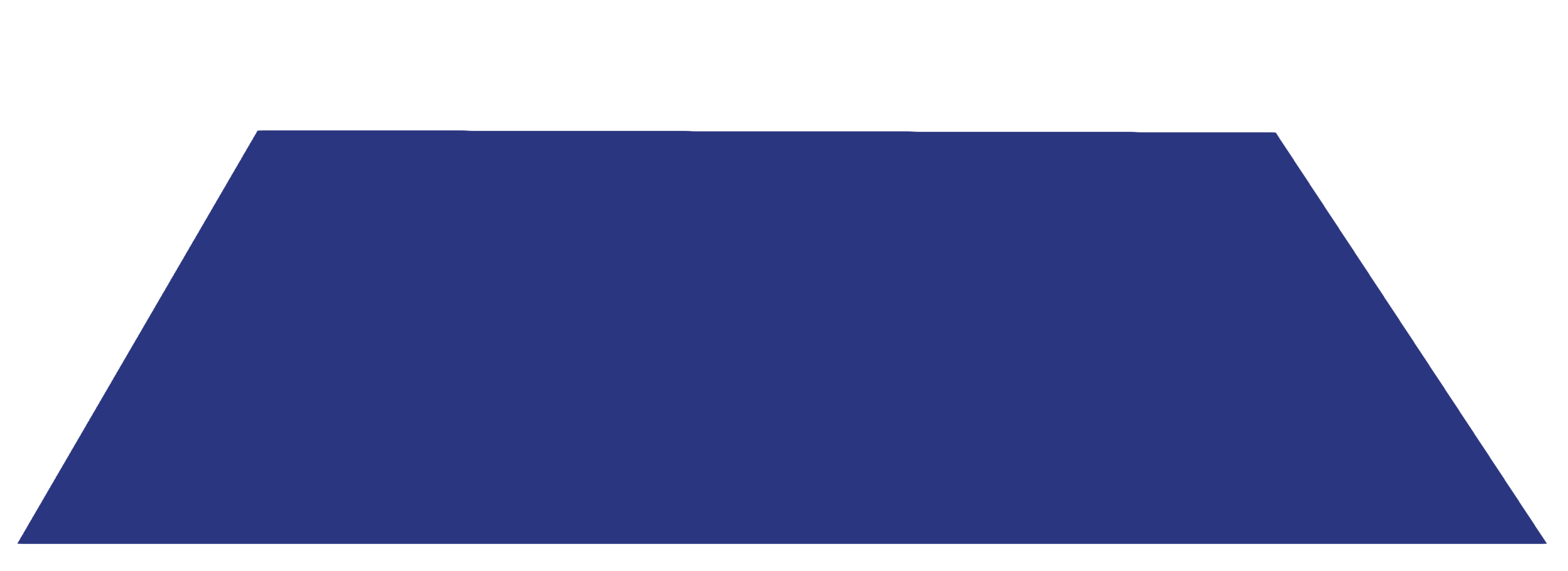}
    \caption{with local correction}
  \end{subfigure}
  \begin{subfigure}[t]{.5\textwidth}
    \centering
    \includegraphics[width=0.8\linewidth]{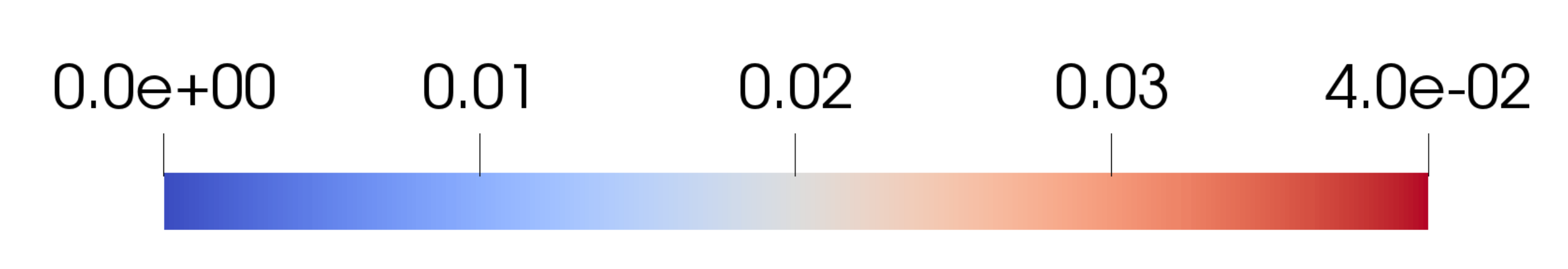}
  \end{subfigure}
  \caption{Absolute value of the error after five multigrid cycles on
  the low quality unit square fine grid (a) without local correction and
  (b) with local correction.}
  \label{fig:square1_res5}
\end{figure}

We test the homogeneous Poisson problem in \cref{sec:model} on $\Omega =
(0, 1)^{2}$ with $f = 0$, $\Gamma_{D} = \Gamma$ and $g = 0$ using linear
elements ($P_1$) and a two-level V-cycle. An initial guess $u^{(0)} =
\sin(10 \pi x) \sin(10\pi y)$ is interpolated onto the fine grid (see
\cref{fig:initial}). The fine grid has $158$ vertices with $272$ cells
and is shown in \cref{fig:squaremesh1} after perturbation of some
vertices. The coarse grid has $68$ cells. One iteration of the smoother
(symmetric Gauss--Seidel as the global smoother) is applied in the pre-
and post-smoothing steps. The norm of the relative residual after each
multigrid cycle is shown in \cref{fig:res_square1} for (i) a low quality
mesh with the standard smoother, (ii) a low quality mesh with the local
correction smoother and (iii) a high quality mesh (reference case).
Slow convergence in the residual with the low quality mesh is clear,
whereas the local correction smoother recovers the rate of the reference
case. \Cref{fig:square1_res5} shows the absolute value of the solution
error on the low quality fine grid (Case~A) after five multigrid
V-cycles, with and without the local correction.
It is clear that the error persists in a localised region around the low
quality cells without the local correction, and this error is
removed by the local correction smoother.

\begin{figure}
  \centering
  \includegraphics[width=0.6\linewidth]{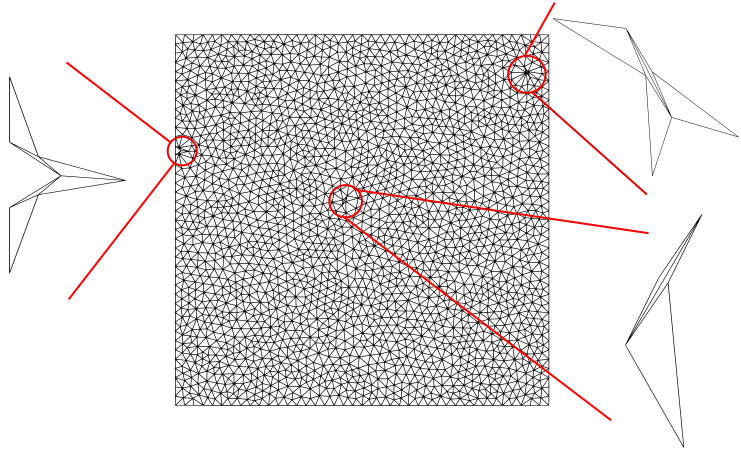}
  \caption{A unit square mesh with three regions of low cell quality.}
  \label{fig:domain_square2}
\end{figure}

\begin{figure}
  \centering
  \begin{subfigure}{.5\textwidth}
    \centering
    \includegraphics[width=0.9\linewidth]{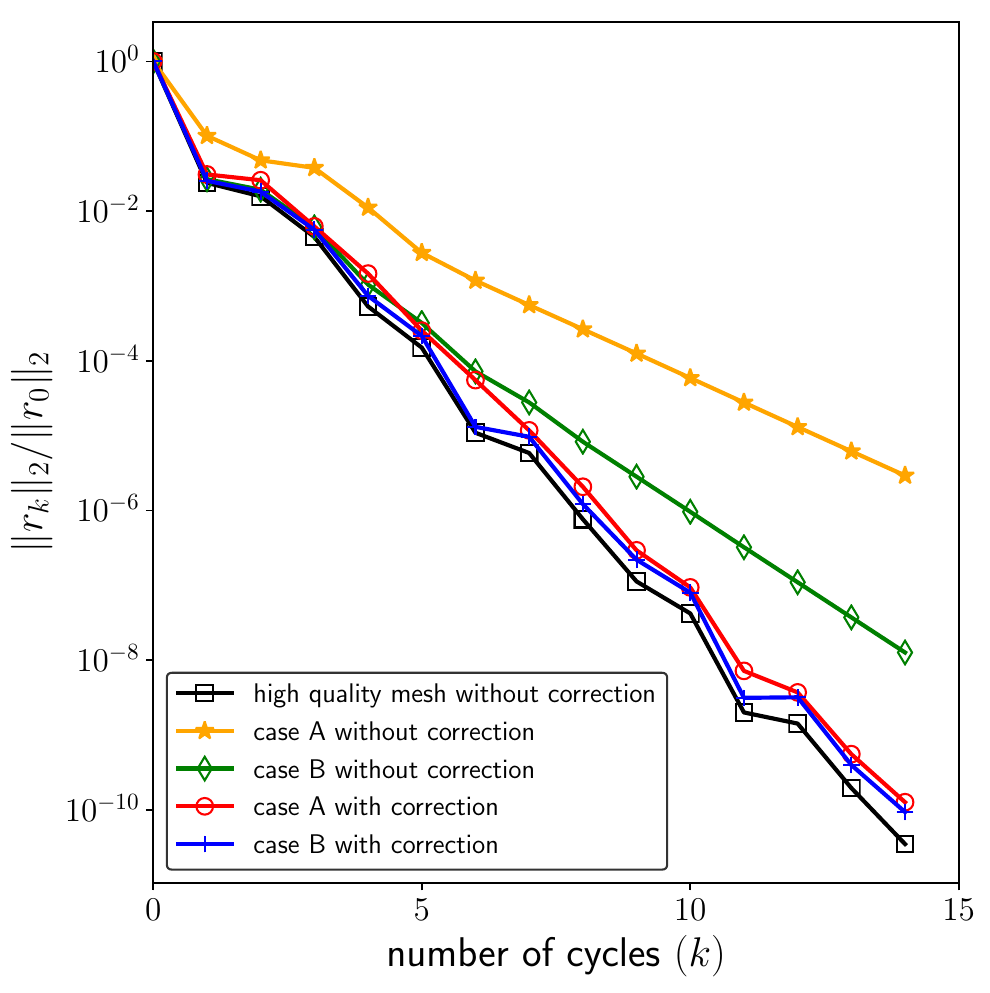}
    \caption{$P_1$ elements}
  \end{subfigure}%
  \begin{subfigure}{.5\textwidth}
    \centering
    \includegraphics[width=0.9\linewidth]{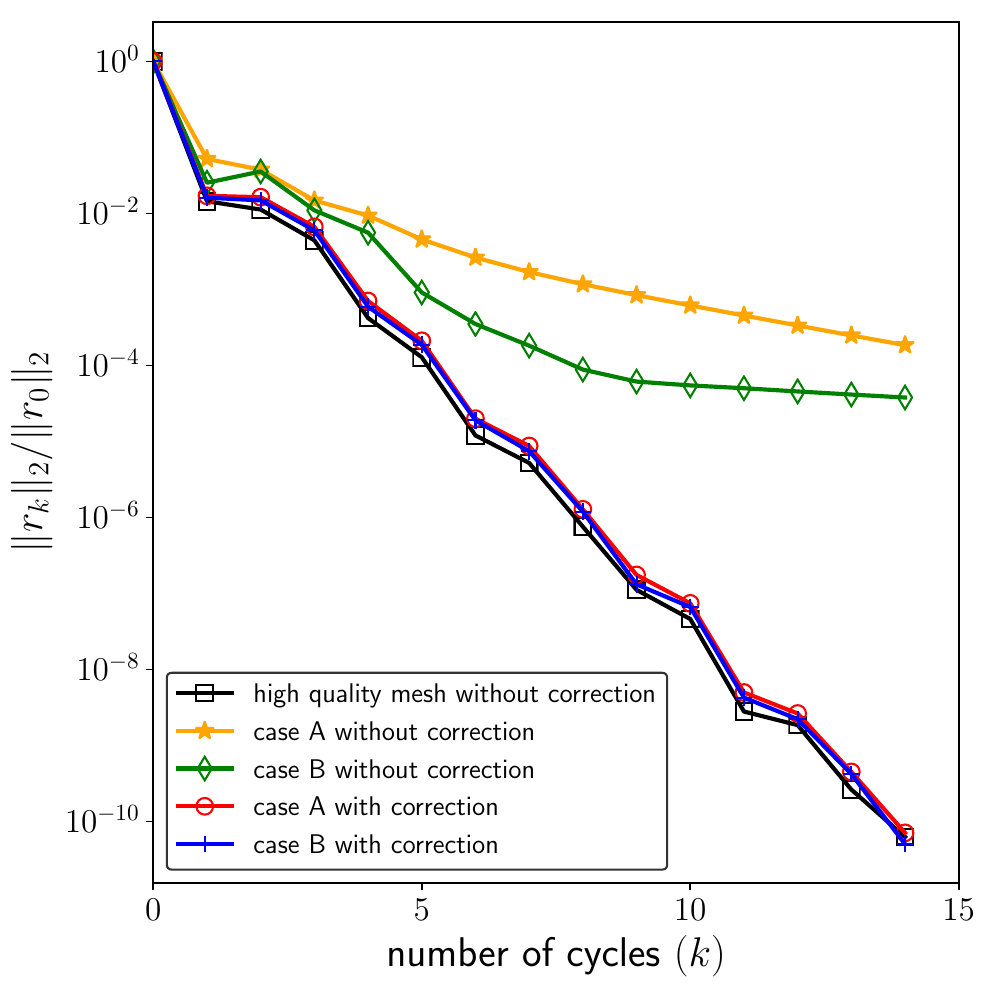}
    \caption{$P_2$ elements}
  \end{subfigure}
  \caption{Relative residual after each multigrid cycle (symmetric
  Gauss--Seidel as global smoother) for the Poisson
  problem on a unit square domain with low quality regions on all levels
  (Case~A) and with low quality regions on all levels except the finest
  level (Case~B).}
  \label{fig:res_square2}
\end{figure}

\begin{figure}
  \centering
  \begin{subfigure}[t]{0.5\textwidth}
    \centering
    \includegraphics[width=0.7\linewidth]{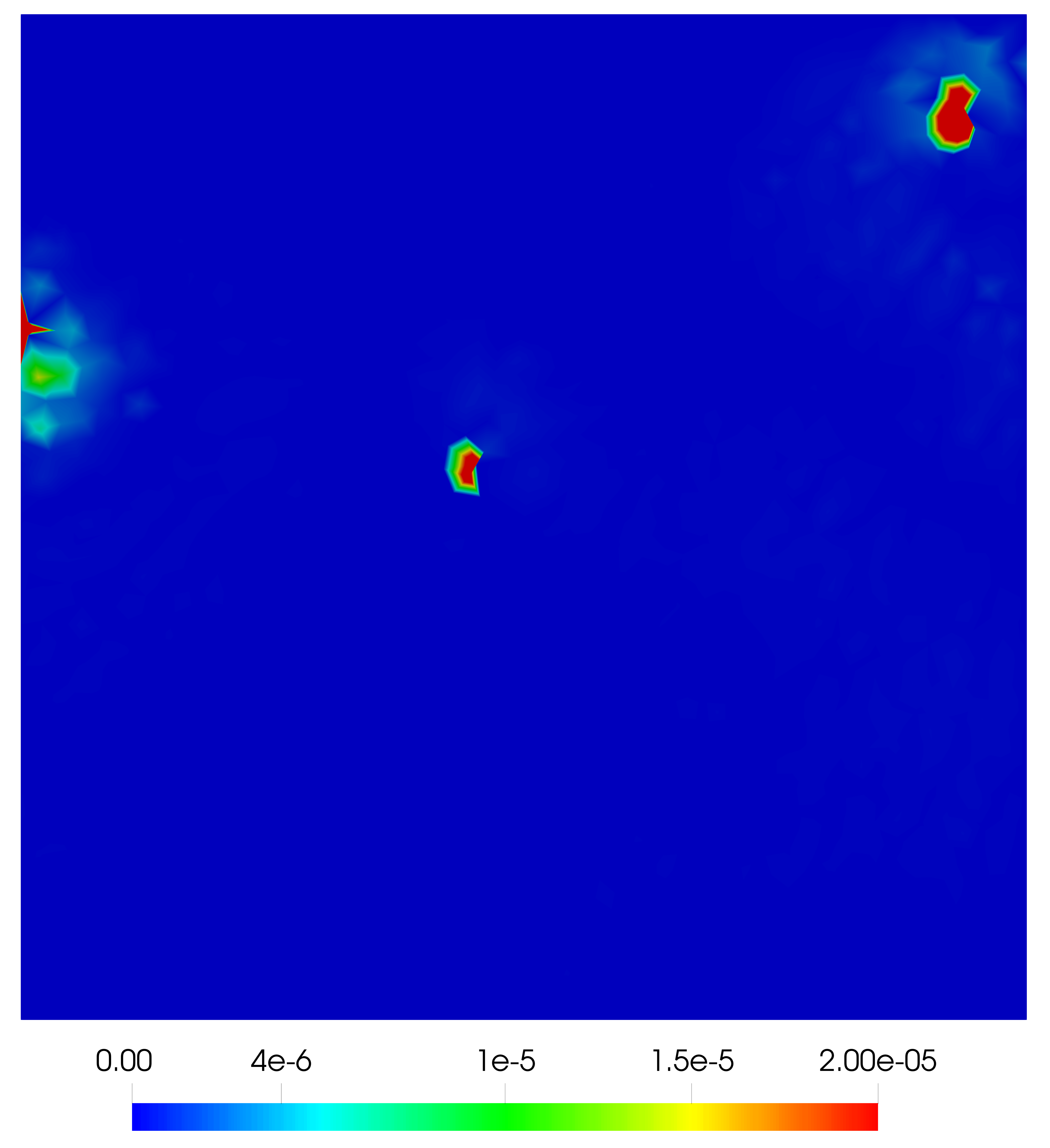}
    \caption{$P_1$ element}
  \end{subfigure}\hfill%
  \begin{subfigure}[t]{0.5\textwidth}
    \centering
    \includegraphics[width=0.7\linewidth]{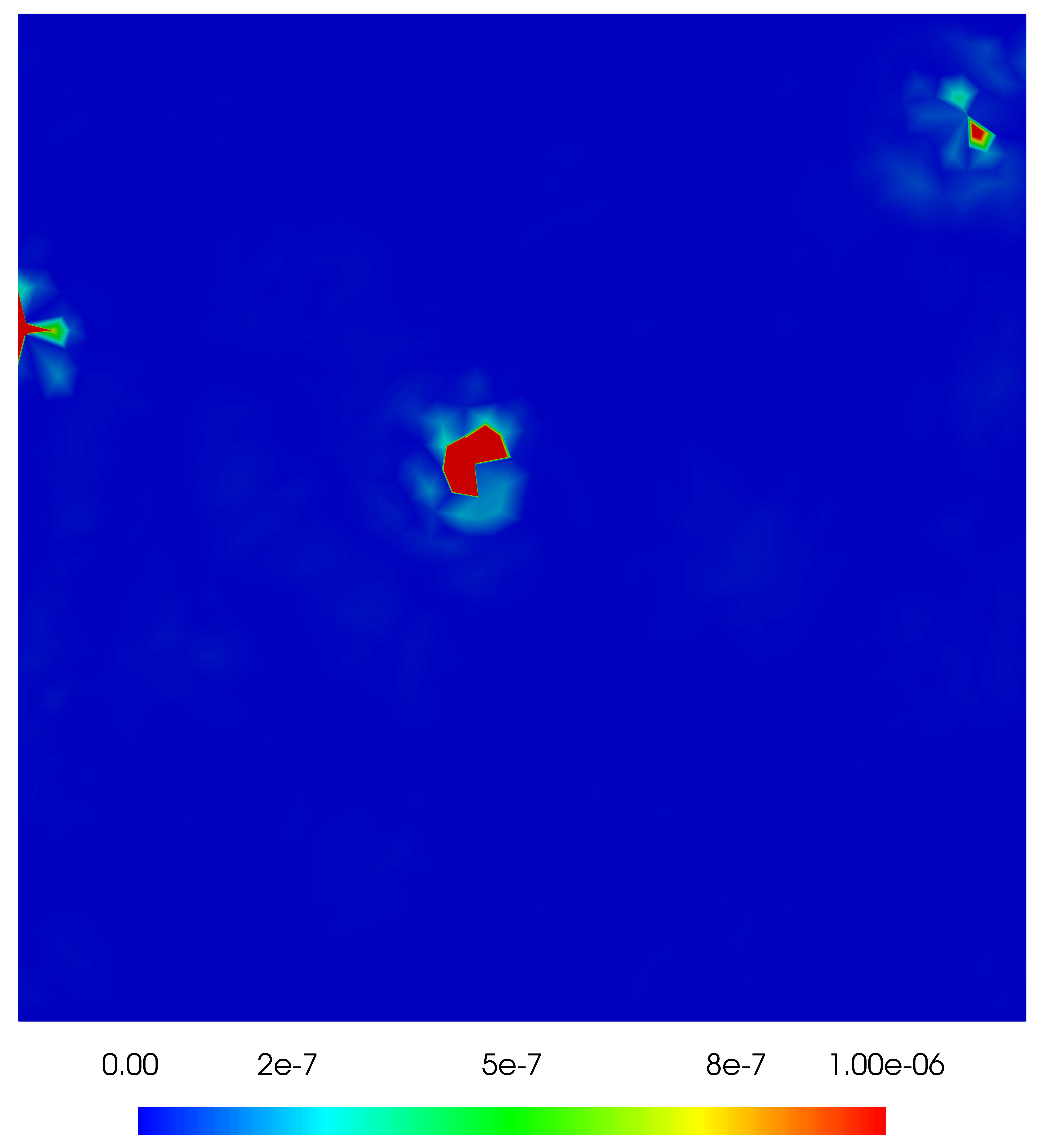}
    \caption{$P_2$ element}
  \end{subfigure}
  \caption{Absolute value of residual at each vertex of the low quality
  finest unit square grid after ten multigrid cycles, without local
  correction, for the four-level Poisson problem using $P_1$ and $P_2$
  elements.}
  \label{fig:contour_square2}
\end{figure}

We next consider a four-level V-cycle, with symmetric Gauss--Seidel as
the standard global smoother. The smoother is applied twice in the pre-
and post-smoothing steps. We perturb the position of some vertices on
each level to generate the low quality meshes (levels). The low quality
finest grid is shown in \cref{fig:domain_square2}, with the regions of low
cell quality indicated.
The number of cells for each level and the number of DOFs for $P_1$ and
$P_2$ elements are summarised in \cref{table:size_square2}, as well as
the cell quality measures and sizes of the degraded regions.
\begin{table}
  \centering
  \begin{tabular}{l|cc|cc|cc}
  \hline
  level
  & $\gamma_{\min}(\Omega_{\text{high}})$
  & $\gamma_{\min}(\Omega_{\text{low}})$
  & \begin{tabular}[c]{@{}c@{}}number of\\ cells in $\Omega$ \end{tabular}
  & \begin{tabular}[c]{@{}c@{}}number of \\ cells in $\Omega_B$ \end{tabular}
  & \begin{tabular}[c]{@{}c@{}}number of \\ DOFs in $\Omega$\\ $P_1/P_2$\end{tabular}
  & \begin{tabular}[c]{@{}c@{}}number of \\ DOFs in $\Omega_B$\\ $P_1/P_2$ \end{tabular} \\
  \hline
  $1$ (fine)   & $0.661$  & $1.53 \times 10^{-4}$  &  \num{4236}   & \num{49}
  & \num{2199}/\num{8633}  & \num{44}/\num{134}  \\
  $2$          & $0.673$  & $3.10 \times 10^{-4}$  &  \num{1016}   & \num{41}
  & \num{549}/\num{2133}   & \num{37}/\num{112}   \\
  $3$          & $0.772$  & $2.09 \times 10^{-4}$  &  \num{254}  & \num{36}
  & \num{148}/\num{549}  & \num{34}/\num{100}   \\
  $4$ (coarse) & $0.773$  & $9.08 \times 10^{-4}$   & \num{68}    & \num{11}
  & \num{45}/\num{157}   & \num{10}/\num{30}    \\
  \hline
  \end{tabular}
  \caption{Cell quality of minimum normalized radius ratio $\gamma$ and
  the problem size on each level of the four--level unit square
  hierarchy meshes.}
  \label{table:size_square2}
\end{table}
We solve the Poisson problem on $\Omega = (0, 1)^2$ with $\Gamma_D =
\{(x, y) \in \partial \Omega: y = 0, 1\}$, $f = 2 \pi^2 \cos(\pi
x)\sin(\pi y) $, $g = 0$ and $s = 0$. \Cref{fig:res_square2} presents
the computed relative residual for Case~A and Case~B, with and without
local correction, for $P_1$ and $P_2$ elements. It is clear that the
convergence rate is slow for low quality meshes, and particularly so for
quadratic elements. In all cases the local correction smoother recovers
the convergence rate of the reference case with high quality meshes.
We consider the error in the solution vector for Case~A, Case~B and the
reference case after 12 multigrid cycles, with the error given by
$\lVert u_{\rm MG} - u_{\rm LU} \rVert_{\ell^2}$, where $u_{\rm MG}$ is
the multigrid solution vector and $u_{\rm LU}$ is the solution vector
computed using LU factorisation. The computed discrete error is
presented in \cref{table:error_square2}.
  \begin{table}
  \centering
  \begin{tabular}{l|cccc}
  \hline
  solver error    & \multicolumn{4}{c}{$\lVert u_{\rm MG} - u_{\rm LU} \rVert_{\ell^2}$}  \\
  \hline
  element type    & \multicolumn{2}{c|}{$P_1$}     & \multicolumn{2}{c}{$P_2$} \\
  \hline
  reference case  & \multicolumn{2}{c|}{$2.86 \times 10^{-8}$}    & \multicolumn{2}{c}{$7.04 \times 10^{-9}$}  \\
  \hline
  case           & Case A & \multicolumn{1}{c|}{Case B} &  Case A & \multicolumn{1}{c}{Case B} \\
  \hline
  without local correction & $2.09 \times 10^{-6}$  & \multicolumn{1}{c|}{$1.57 \times 10^{-7}$}
  & $3.91 \times 10^{-5}$     &   $3.76 \times 10^{-5}$ \\
  \hline
  with local correction    & $8.04 \times 10^{-8}$    & \multicolumn{1}{c|}{$6.35 \times 10^{-8}$}
  & $1.10\times 10^{-8}$   & $1.01 \times 10^{-8}$   \\
  \hline
  \end{tabular}
  \caption{Solution vector error in the $\ell^{2}$ norm after $12$
   multigrid cycles with and without local correction for the Poisson
   problem on a unit square domain.}
  \label{table:error_square2}
  \end{table}
With the proposed smoother, the solver error is reduced, particularly
for $P_2$ elements. The finite element solution error in the
$\mathcal{L}^2$ norm for $P_2$ elements is shown in
\cref{table:fem_error_square2} after different numbers of cycles.
\begin{table}
  \centering
  \begin{tabular}{cccc}
  \multicolumn{4}{c}{FEM solution error $\lVert u_{h} - u \rVert_{\mathcal{L}^2(\Omega)}$} \\
  \hline
  \multicolumn{1}{c|}{\begin{tabular}[c]{@{}c@{}}number of \\cycles ($k$) \end{tabular}}
  &\multicolumn{1}{c|}{reference case}
  & \multicolumn{1}{c|}{\begin{tabular}[c]{@{}c@{}}Case A\\ without local correction\end{tabular}}
  & \begin{tabular}[c]{@{}c@{}}Case A\\ with local correction\end{tabular} \\
  \hline
  \multicolumn{1}{c|}{2}   & \multicolumn{1}{c|}{$1.28 \times 10^{-3}$} & \multicolumn{1}{c|}{$2.18\times 10^{-3}$}  & $1.46\times 10^{-3}$   \\
  \multicolumn{1}{c|}{5}   & \multicolumn{1}{c|}{$6.05 \times 10^{-6}$} & \multicolumn{1}{c|}{$3.67\times 10^{-5}$}  & $1.04\times 10^{-5}$   \\
  \multicolumn{1}{c|}{10}  & \multicolumn{1}{c|}{$2.56\times 10^{-6}$} & \multicolumn{1}{c|}{$3.83\times 10^{-6}$}  & $2.70\times 10^{-6}$    \\
  \hline \hline
  \multicolumn{1}{c|}{\begin{tabular}[c]{@{}c@{}}number of \\cycles ($k$) \end{tabular}}
  &\multicolumn{1}{c|}{reference case}
  & \multicolumn{1}{c|}{\begin{tabular}[c]{@{}c@{}}Case B\\ without local correction\end{tabular}}
  & \begin{tabular}[c]{@{}c@{}}Case B\\ with local correction\end{tabular} \\
  \hline
  \multicolumn{1}{c|}{2}  & \multicolumn{1}{c|}{$1.28 \times 10^{-3}$} & \multicolumn{1}{c|}{$2.11\times 10^{-3}$}  & $1.37\times 10^{-3}$ \\
  \multicolumn{1}{c|}{5}  & \multicolumn{1}{c|}{$6.05 \times 10^{-6}$} & \multicolumn{1}{c|}{$3.64\times 10^{-5}$}  & $9.74\times 10^{-6}$ \\
  \multicolumn{1}{c|}{10}  & \multicolumn{1}{c|}{$2.56\times 10^{-6}$} & \multicolumn{1}{c|}{$3.61\times 10^{-6}$}  & $2.56\times 10^{-6}$  \\
  \hline
  \end{tabular}
  \caption{Finite element solution error in the $\mathcal{L}^2$-norm for
  solving the Poisson equation on a unit square domain with and without
  local correction for $P_2$ elements, where $u_{h}$ is the computed
  solution (after a specified number of multigrid cycles) and $u$ is the
  exact solution to the Poisson problem.}
  \label{table:fem_error_square2}
\end{table}
The error is greatest for low quality meshes without the local
correction. With the local correction, the finite element solution error
is reduced. For Case~B, which has a high quality fine grid, the accuracy
with the local correction is the same as the reference case after ten
cycles. Low quality intermediate grids have not degraded accuracy and
the local correction overcomes slow solver convergence.
\Cref{fig:contour_square2} shows the absolute value of discrete residual
on the finest grid after 10 V-cycles for the case where low quality
cells appear on all levels and local correction is only applied on
coarse grids. Unsurprisingly, the large value residual persists in the
regions with low quality cells.

\subsubsection{Unit cube domain}

\begin{figure}
  \centering
  \begin{subfigure}[t]{0.5\textwidth}
    \centering
    \includegraphics[width=0.8\linewidth]{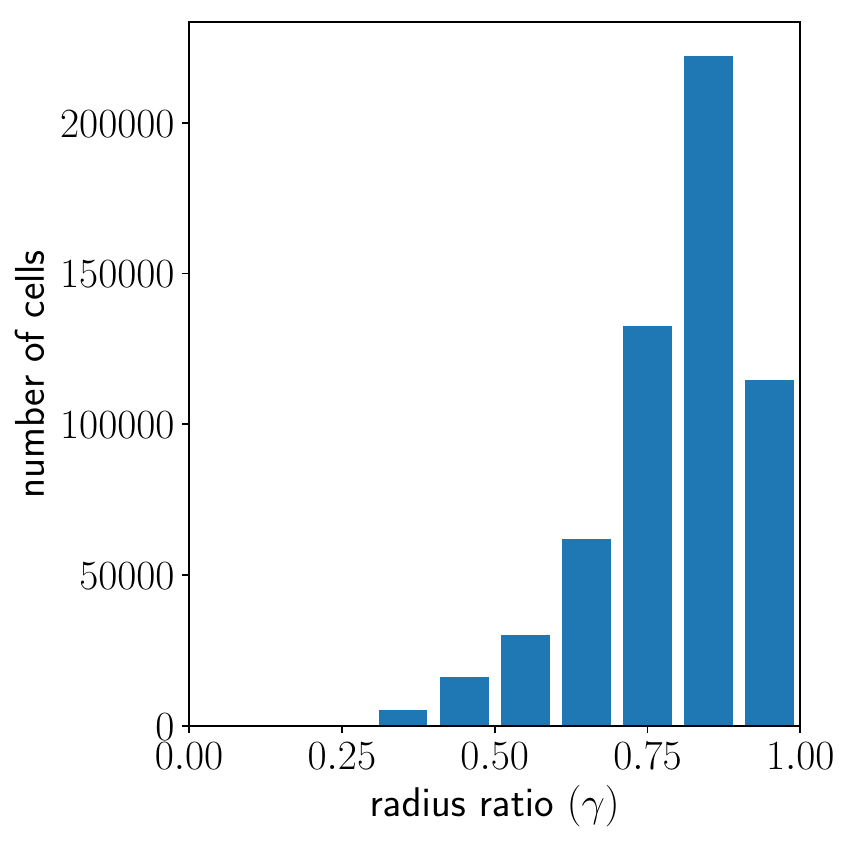}
    \caption{Full range of normalised radius ratio.}
  \end{subfigure}%
  \begin{subfigure}[t]{0.5\textwidth}
    \centering
    \includegraphics[width=0.8\linewidth]{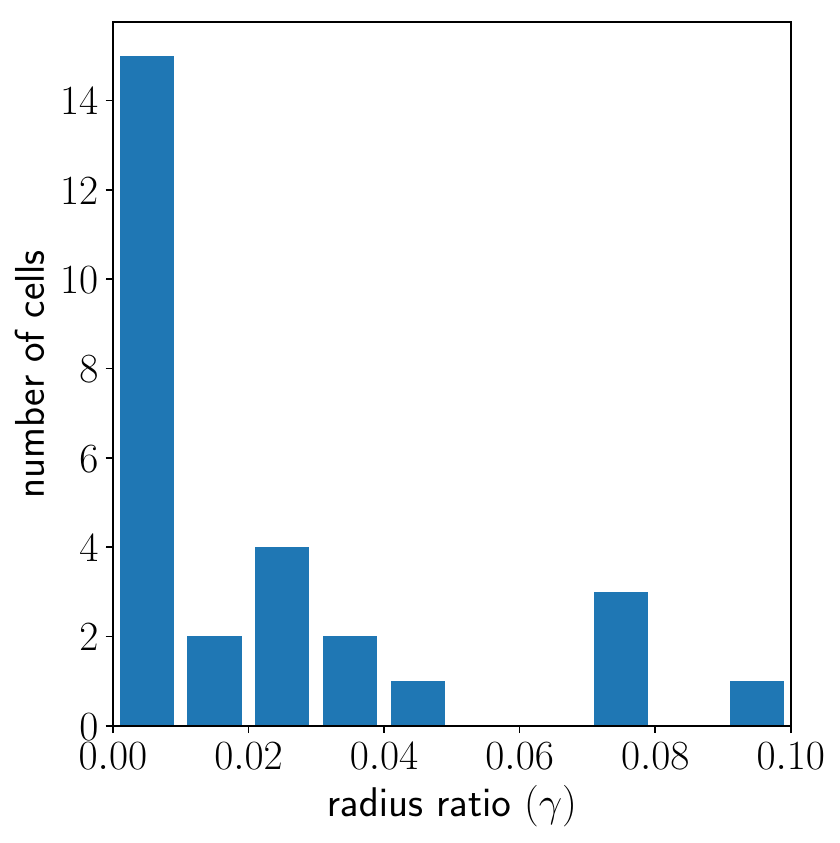}
    \caption{Number of cells with a normalised radius ratio of less
    than~$0.1$.}
  \end{subfigure}
  \caption{Histogram of the cell normalised radius ratio $\gamma$
  (\cref{eqn:radius_ratio}) for a unit cube fine grid with low quality
  regions.}
  \label{fig:hist_cube}
\end{figure}

We consider $\Omega = (0, 1)^3$ with $\Gamma_{D} = \{(x, y, z) \in
\partial \Omega: x=0, 1\}$, $f = 10\exp[-((x - 0.5)^2 + (y - 0.5)^2 +
(z-0.5)^2) / 0.02]$, $g = 0$ and $s = 0$. We use a four-level multigrid
V-cycle for this problem. The coarsening rate in terms of number of
degrees-of-freedom is in the range of~7--8 at each level. The cell
quality measures and cell sizes for all grid levels
($\Omega_{\text{high}}$ and $\Omega_{\text{low}}$) are summarised in
\cref{table:size_cube}. A histogram of the normalised radius ratio for
the Case~A (low quality finest grid) is shown in \cref{fig:hist_cube}.
\begin{table}
  \footnotesize
  \centering
  \begin{tabular}{l|cc|cc|cc}
  \hline
  level
  & $\gamma_{\min}(\Omega_{\text{high}})$
  & $\gamma_{\min}(\Omega_{\text{low}})$
  & \begin{tabular}[c]{@{}c@{}}number of\\ cells in $\Omega$ \end{tabular}
  & \begin{tabular}[c]{@{}c@{}}number of \\ cells in $\Omega_B$ \end{tabular}
  & \begin{tabular}[c]{@{}c@{}}number of \\ DOFs in $\Omega$\\ $P_1/P_2$\end{tabular}
  & \begin{tabular}[c]{@{}c@{}}number of \\ DOFs in $\Omega_B$\\ $P_1/P_2$ \end{tabular} \\
  \hline
  1 (fine) & $0.275$ & $7.20\times10^{-6}$ & \num{582730}   & \num{625}
    & \num{104976}/\num{814775}  & \num{231}/\num{1262} \\
  2  & $0.288$ & $2.77\times10^{-6}$  & \num{65259}   & \num{604}
    & \num{13361}/\num{97422}     & \num{244}/\num{1294} \\
  3  & $0.278$ & $1.74\times10^{-6}$ & \num{7165}   & \num{568}
    & \num{1776}/\num{11845}      & \num{233}/\num{1226} \\
  4 (coarse) & $0.336$  & $4.64\times10^{-6}$  & \num{792}   & \num{116}
    & \num{251}/\num{1501}     & \num{50}/\num{258}\\
      \hline
  \end{tabular}
  \caption{Cell quality of minimum normalized radius ratio $\gamma$ and
  the problem size on each level of the unit cube hierarchy of meshes.}
  \label{table:size_cube}
\end{table}

\begin{figure}
  \centering
  \begin{subfigure}{0.5\textwidth}
    \centering
    \includegraphics[width=0.9\linewidth]{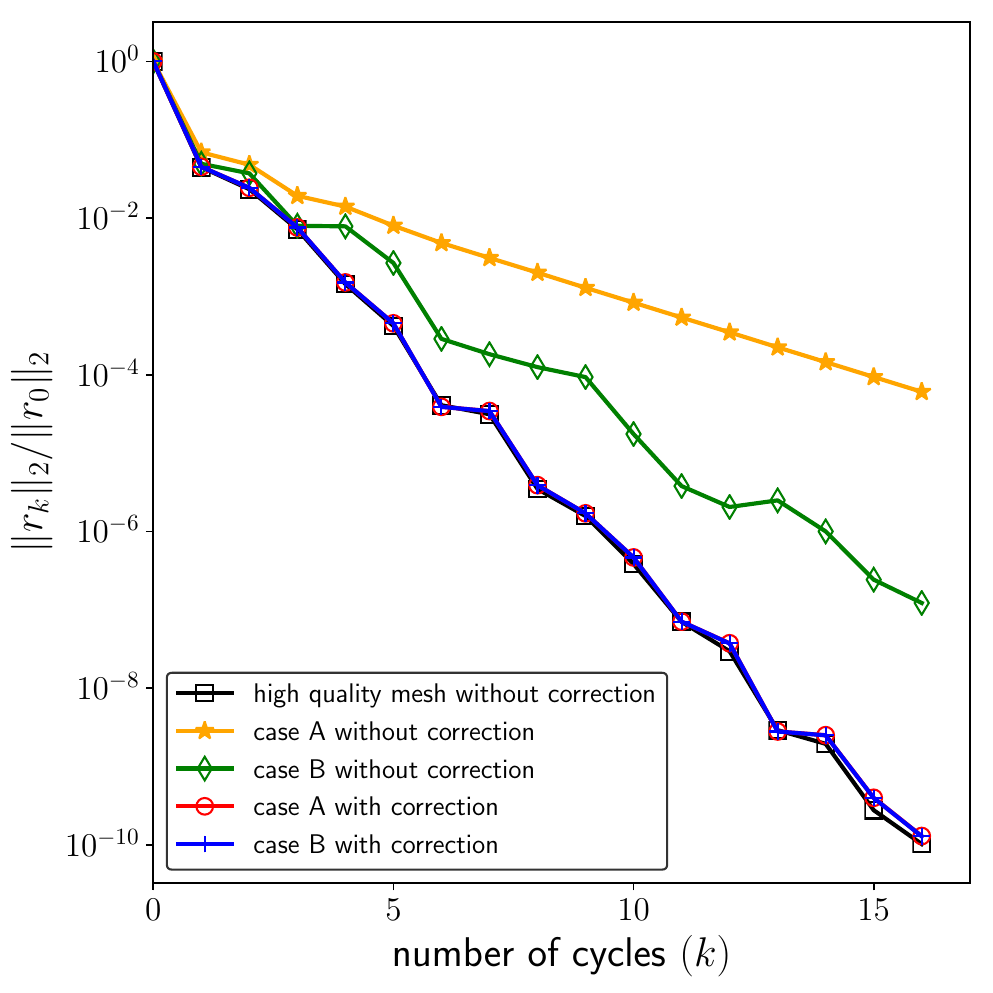}
    \caption{$P_1$ element}
  \end{subfigure}%
  \begin{subfigure}{0.5\textwidth}
    \centering
    \includegraphics[width=0.9\linewidth]{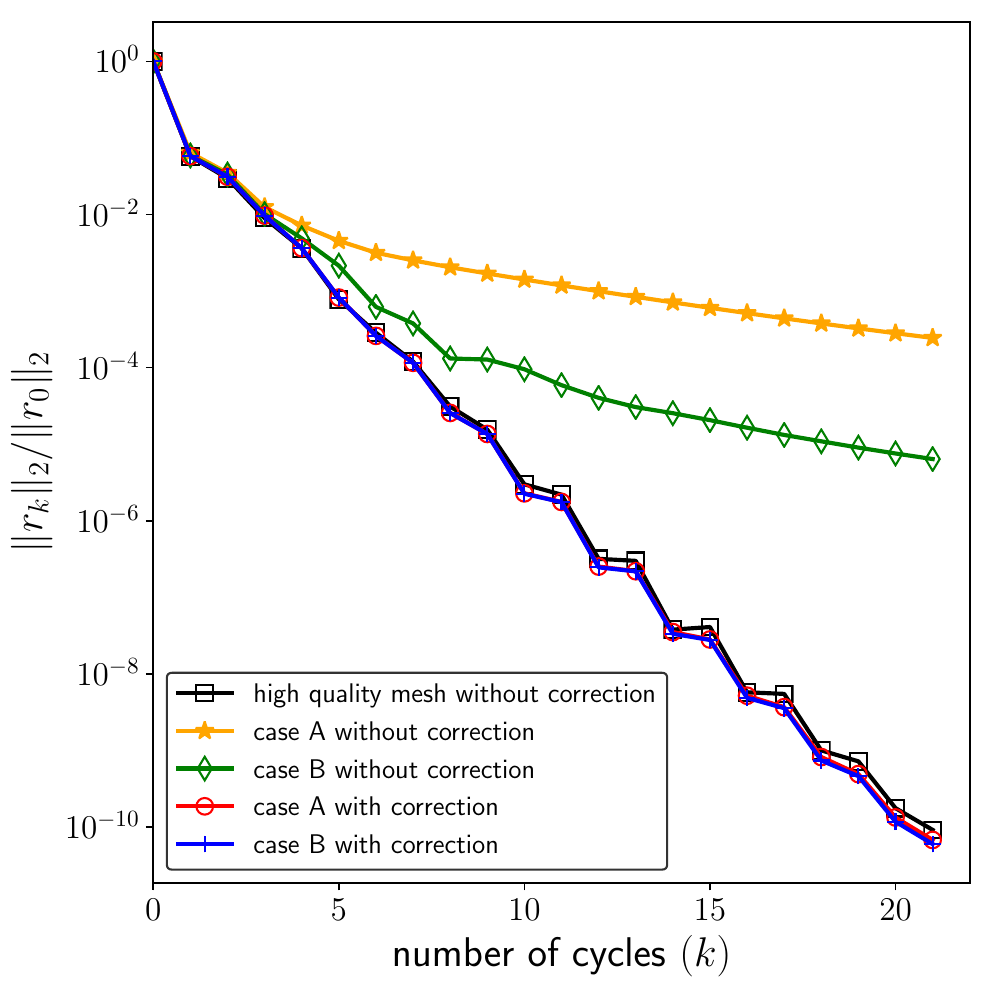}
    \caption{$P_2$ element}
  \end{subfigure}
  \caption{Relative residual after each multigrid cycle (symmetric
  Gauss--Seidel as global smoother) for for the Poisson problem on
  the unit cube domain with low quality regions on all
  levels (Case~A) and with low quality regions on all levels except the
  finest level (Case~B).}
  \label{fig:res_cube_gs}
\end{figure}

\Cref{fig:res_cube_gs} presents the relative residual after each
multigrid cycle using symmetric Gauss--Seidel smoother as the global
smoother, with and without the local correction. Two iterations of the
smoother are applied in pre- and in post-smoothing. The reduction
of the residual is again slow with the standard smoother for cases with
a low quality grid level, particularly for $P_2$ elements. The local
correction restores the convergence rate to that of the high quality
mesh reference case.

\begin{figure}
  \centering
  \begin{subfigure}{.5\textwidth}
    \centering
    \includegraphics[width=0.8\linewidth]{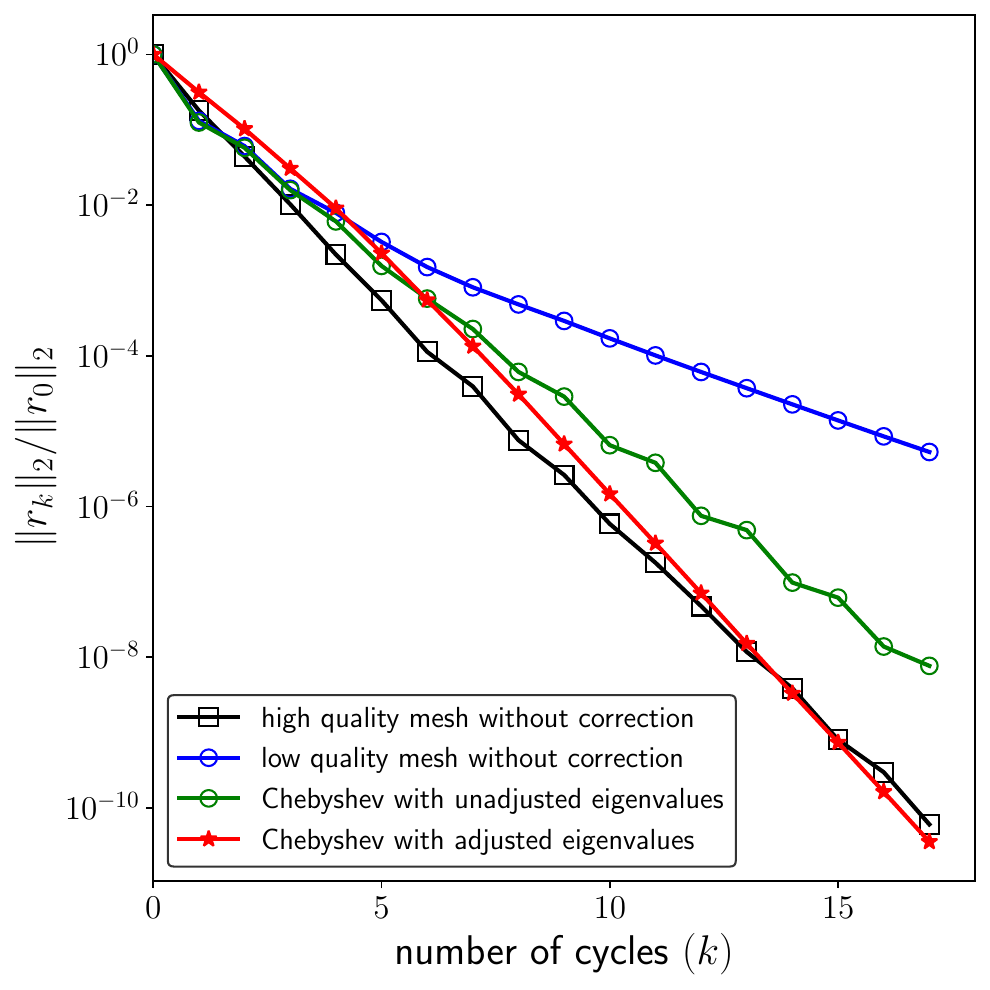}
    \caption{$P_1$ element}
  \end{subfigure}%
  \begin{subfigure}{.5\textwidth}
    \centering
    \includegraphics[width=0.8\linewidth]{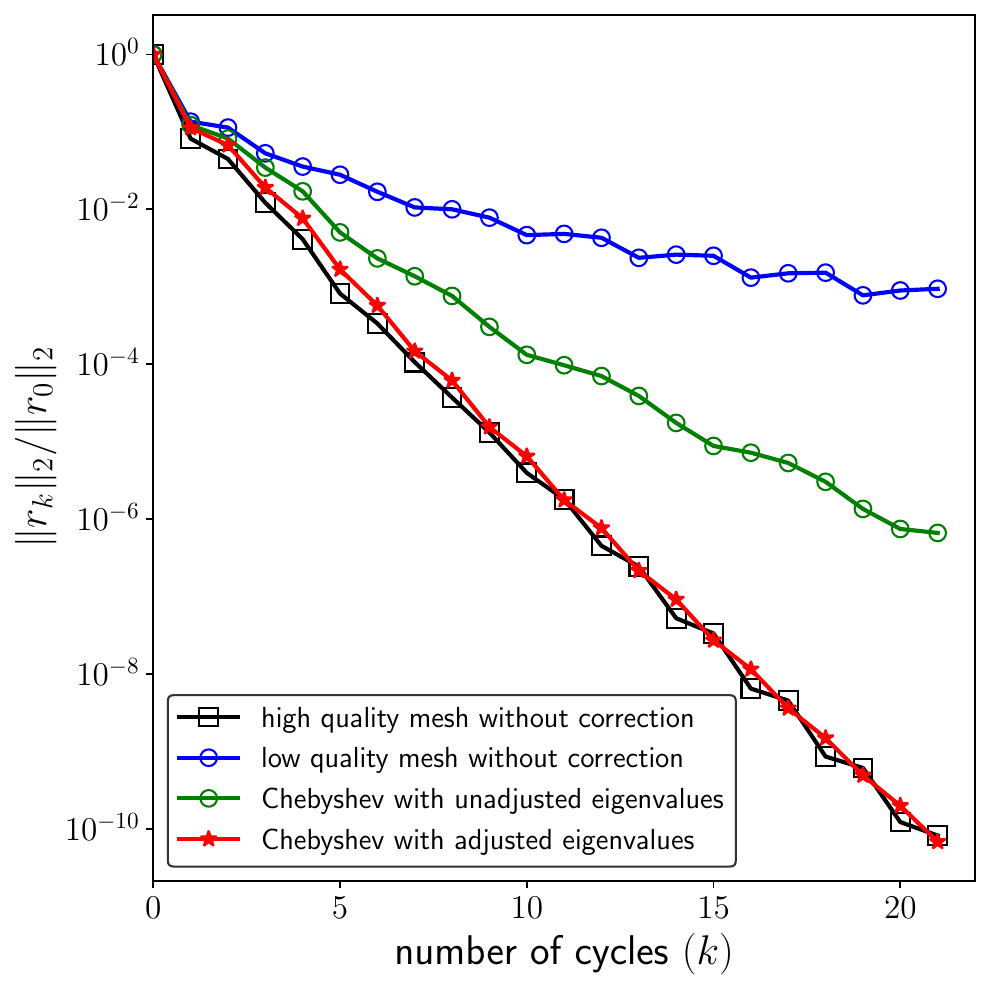}
    \caption{$P_2$ element}
  \end{subfigure}
  \caption{Relative residual after each multigrid cycle (Chebyshev
  global smoother, using maximum eigenvalues (unadjusted) and using
  adjusted eigenvalues) for Poisson problem on the unit cube mesh with
  low quality regions on all levels (Case~A).}
  \label{fig:res_cube_shift}
\end{figure}

We now replace the symmetric Gauss--Seidel smoother with a Jacobi
preconditioned Chebyshev smoother and test performance for Case~A (all
levels contain low quality cells). The smallest eigenvalue is
approximated as one tenth of the largest eigenvalue. We use the (i)
largest eigenvalue (unadjusted) and the (ii) adjusted largest eigenvalue
in the Chebyshev smoother. The largest eigenvalues for high and low
quality meshes, as well as the adjusted largest eigenvalues
(\cref{sec:chebyshev}), are presented in \cref{table:eig_cube}.
\begin{table}
  \centering
  \begin{tabular}{lcccccc}
  \multicolumn{7}{c}{largest eigenvalue}    \\
  \hline
  \multicolumn{1}{l|}{element type}      & \multicolumn{3}{c|}{$P_1$}         & \multicolumn{3}{c}{$P_2$}      \\
  \hline
  \multicolumn{1}{l|}{level}    & level 1 & level 2 & \multicolumn{1}{c|}{level 3}  & level 1 & level 2 & level 3 \\
  \hline
  \multicolumn{1}{l|}{high quality mesh}    & 2.0787  & 2.0227  & \multicolumn{1}{c|}{2.3897}  & 2.4544  & 2.5602  & 2.2711  \\
  \hline
  \multicolumn{1}{l|}{low quality mesh}     & 3.7103  & 2.6002  & \multicolumn{1}{c|}{3.8458}  & 4.4079  & 9.5812  & 10.1333 \\
  \hline
  \multicolumn{1}{l|}{\begin{tabular}[l]{@{}l@{}}adjusted largest eigenvalue\\ for low quality mesh\end{tabular}}
  & 2.0788  & 2.1126  & \multicolumn{1}{c|}{2.3756}  & 2.4543  & 3.4003  & 2.8149  \\
  \hline
  \end{tabular}
  \caption{Largest eigenvalues on high and low quality unit cube meshes,
  and the adjusted largest eigenvalues for low quality unit cube meshes
  (Poisson problem).}
  \label{table:eig_cube}
\end{table}
The residual after each cycle is presented in \cref{fig:res_cube_shift}.
The convergence rate improves using the adjusted eigenvalues in the
Chebyshev smoother, and matches the reference case.

\subsection{Linear elasticity}

We now consider the linearised elasticity problem,
\begin{equation}
  - \nabla \cdot \sigma(u) = f \quad \text{in} \quad \Omega,
\end{equation}
where $u$ is the displacement field and $\sigma(u)$ is the stress tensor
satisfying the isotropic elastic law
\begin{equation}
  \sigma(u) := 2\mu \epsilon(u) + \lambda \text{tr}(\epsilon(u))I,
\end{equation}
where $\epsilon(u)$ is the strain,
\begin{equation}
  \epsilon(u) := \frac{1}{2} \left(\nabla u+ (\nabla u)^T \right),
\end{equation}
and $\mu := E / 2(1 +\nu)$ and $\lambda := E \nu / (1+\nu)(1-2\nu)$ are
the Lam{\'e} parameters, $E$ is the Young's modulus and $\nu$ is
Poisson's ratio. For each example we use $E = 6.9\times 10^{10}$
and~$\nu = 0.33$.

For elasticity examples, multigrid is used as a preconditioner for the
conjugate gradient (CG) method, with one multigrid V-cycle applied at
each CG iteration. In each multigrid cycle, the smoother is applied
twice in pre- and twice in post-smoothing. All examples use four
multigrid levels.

\subsubsection{Lattice domain}

\begin{figure}
  \centering
  \begin{subfigure}{.5\textwidth}
    \centering
    \includegraphics[width=1.0\linewidth]{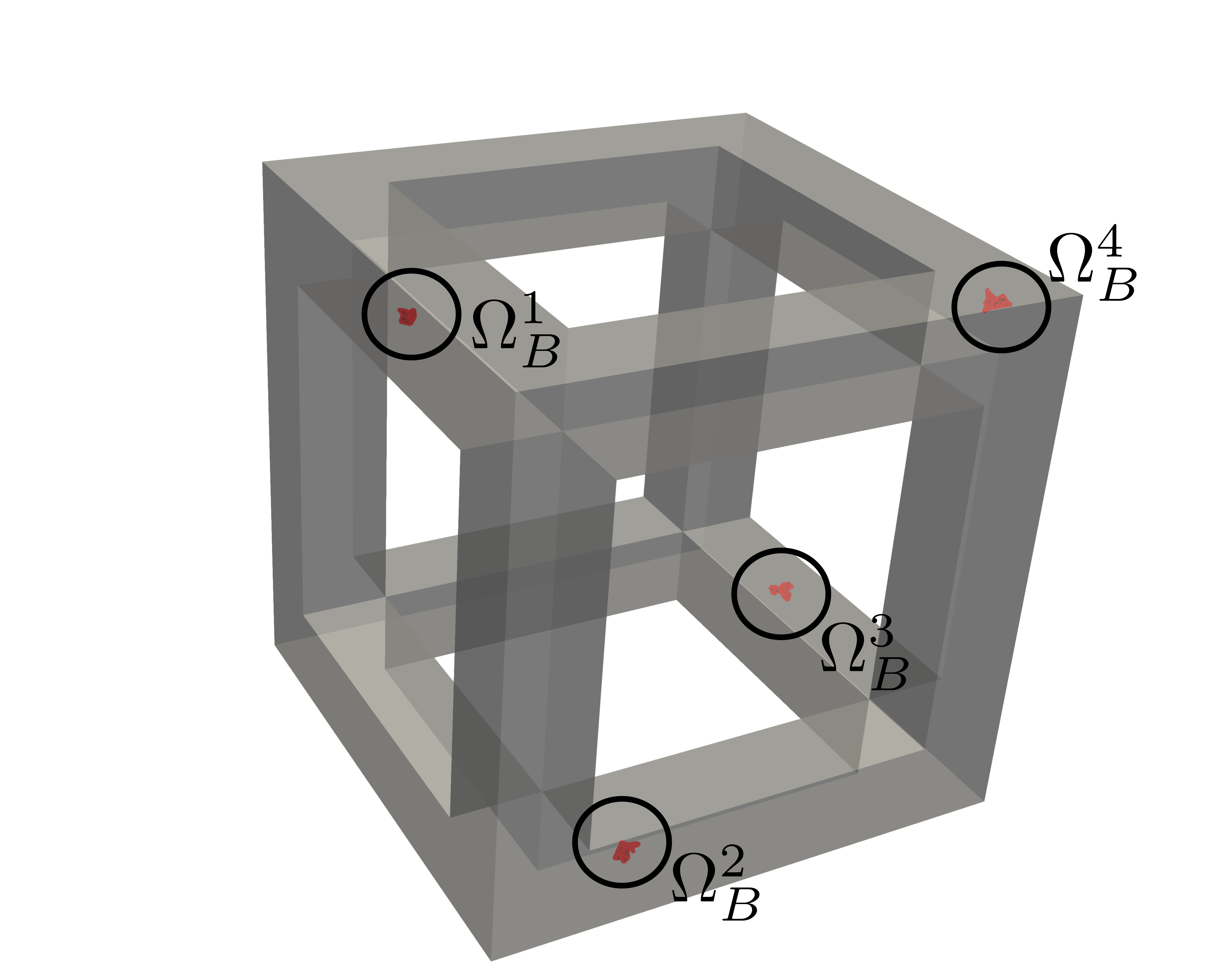}
    \caption{Low quality cells}
  \end{subfigure}%
  \begin{subfigure}{.5\textwidth}
    \centering
    \includegraphics[width=0.7\linewidth]{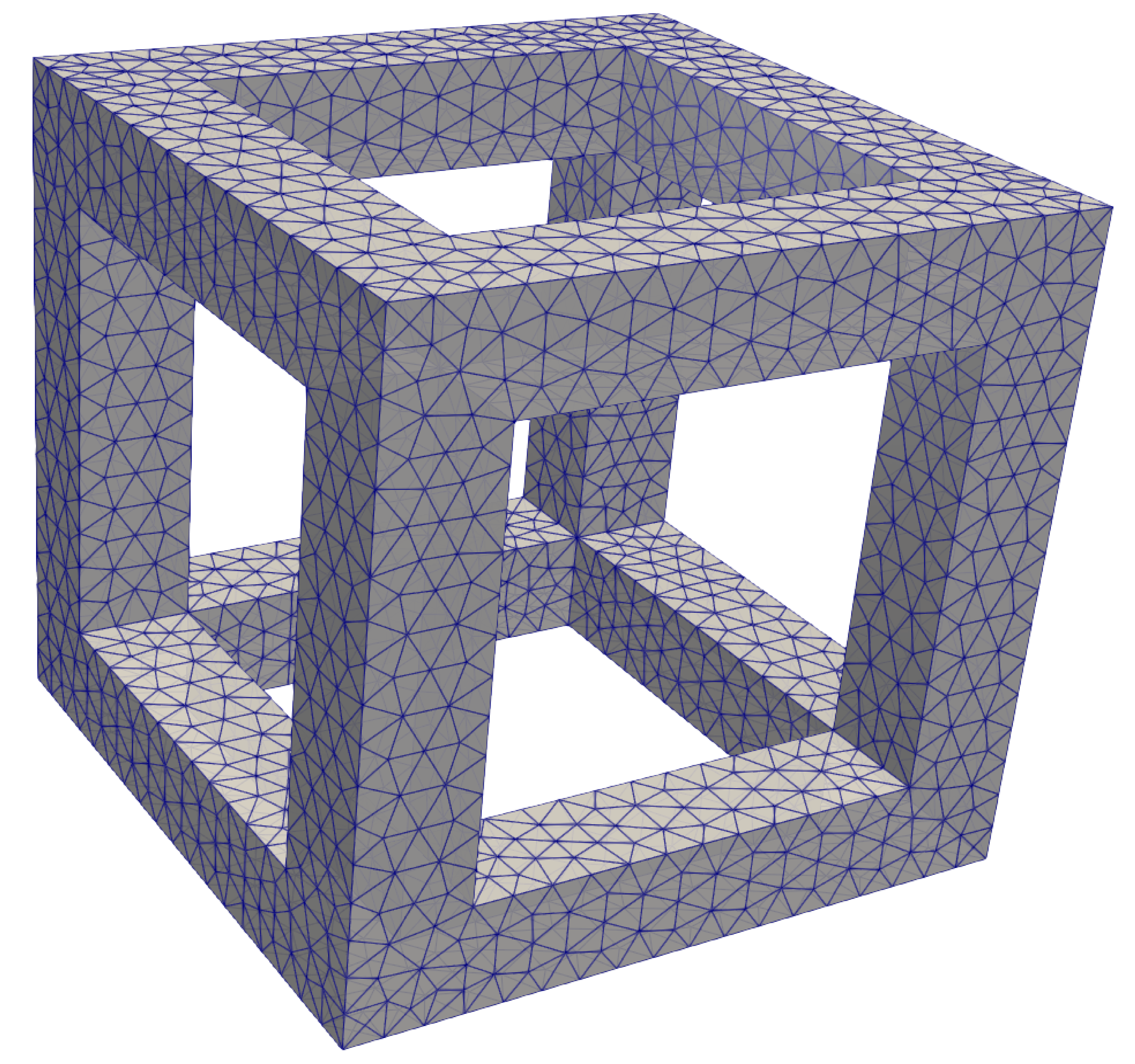}
    \caption{unstructured mesh}
  \end{subfigure}
  \caption{Lattice geometry and the positions of low quality cells on
  the finest grid.}
  \label{fig:domain_lattice}
\end{figure}

The domain tested here is a truss-like lattice structure $\Omega =a^3
\setminus (a\times b^2 \cup b\times a \times b \cup b^2 \times a )$,
with $a = [0, 6]$ and $b = [1, 5]$, which is illustrated in
\cref{fig:domain_lattice}. The locations of low quality regions for the
low quality fine grid are indicated in \cref{fig:domain_lattice}.
We consider $f = (0, 0, 0)$ and boundary conditions
\begin{equation}
  \begin{aligned}
  u = (0,0,0) \quad &\text{on} \quad
  \Gamma_1 = \{(x,y,z)\in \partial \Omega: x=0\},
  \\
  \sigma \cdot n =(10^3,0,0) \quad &\text{on}
  \quad \Gamma_2=\{(x,y,z)\in \partial \Omega: x=6\},
  \\
  \sigma \cdot n = 0   \quad &\text{on}
  \quad \partial \Omega \setminus \{\Gamma_1 \cup \Gamma_2\}.
  \end{aligned}
\end{equation}
Data on the size and cell quality of each grid level for both high and
low quality meshes is summarised in \cref{table:size_lattice}.
\begin{table}
  \footnotesize
  \centering
  \begin{tabular}{l|cc|cc|cc}
  \hline
  level
  & $\gamma_{\min}(\Omega_{\text{high}})$
  & $\gamma_{\min}(\Omega_{\text{low}})$
  & \begin{tabular}[c]{@{}c@{}}number of\\ cells in $\Omega$ \end{tabular}
  & \begin{tabular}[c]{@{}c@{}}number of \\ cells in $\Omega_B$ \end{tabular}
  & \begin{tabular}[c]{@{}c@{}}number of \\ DOFs in $\Omega$\\ $P_1/P_2$\end{tabular}
  & \begin{tabular}[c]{@{}c@{}}number of \\ DOFs in $\Omega_B$\\ $P_1/P_2$ \end{tabular} \\
  \hline
  1 (fine) & $0.209$ & $2.64\times 10^{-8}$ & \num{711,683}  & \num{732}
      & \num{437946}/\num{3192258}  & \num{825}/\num{4506} \\
  2  & $0.228$ & $2.70\times 10^{-8}$ & \num{78100}    & \num{585}
      & \num{58992}/\num{391767}     & \num{798}/\num{4101} \\
  3  & $0.250$ & $6.43\times 10^{-8}$ & \num{8341}   & \num{444}
      & \num{8292}/\num{49131}      & \num{612}/\num{3027} \\
  4 (coarse) & $0.344$  & $1.07\times 10^{-6}$  & \num{1080}    & \num{96}
      & \num{1284}/\num{7116}     & \num{183}/\num{816}\\
  \hline
  \end{tabular}
  \caption{Cell quality of minimum normalized radius ratio $\gamma$ and
    the problem size on each level of the lattice hierarchy meshes.}
  \label{table:size_lattice}
\end{table}

\begin{figure}
  \centering
  \begin{subfigure}{.5\textwidth}
    \centering
    \includegraphics[width=0.9\linewidth]{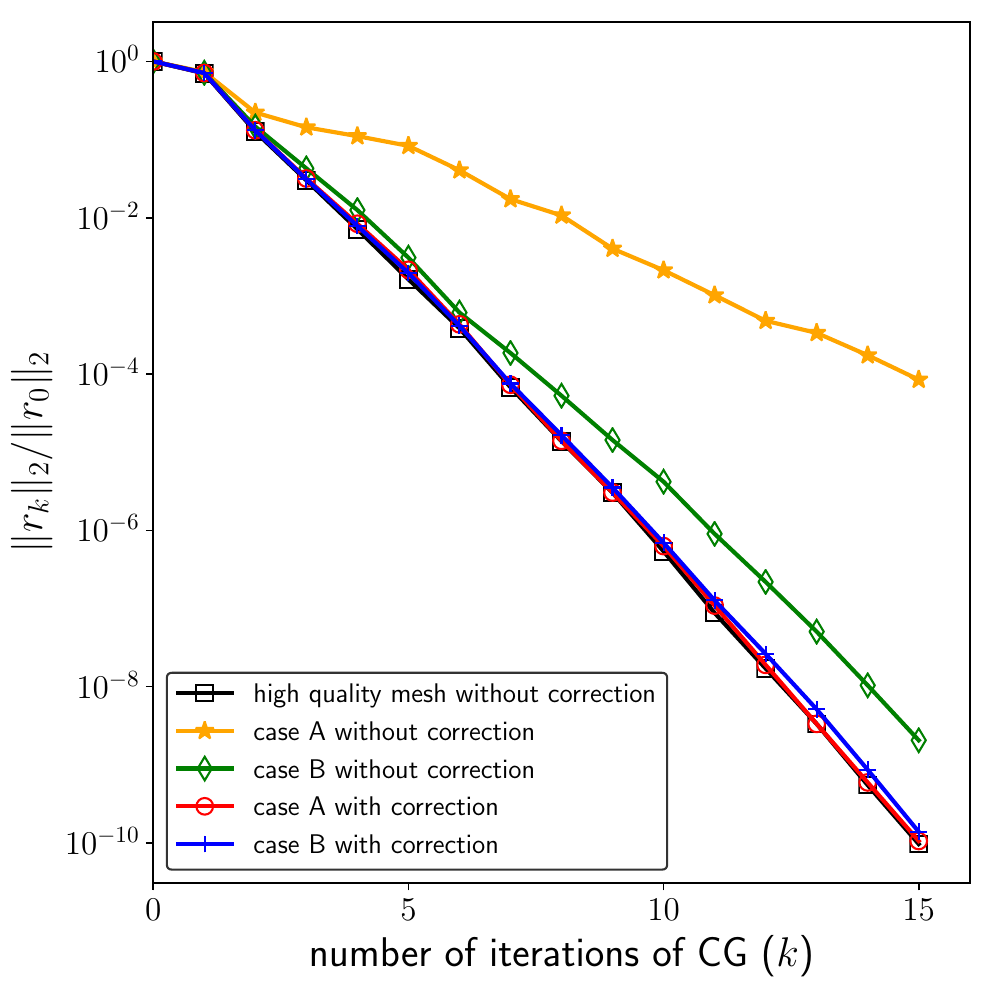}
    \caption{$P_1$ element}
  \end{subfigure}%
  \begin{subfigure}{.5\textwidth}
    \centering
    \includegraphics[width=0.9\linewidth]{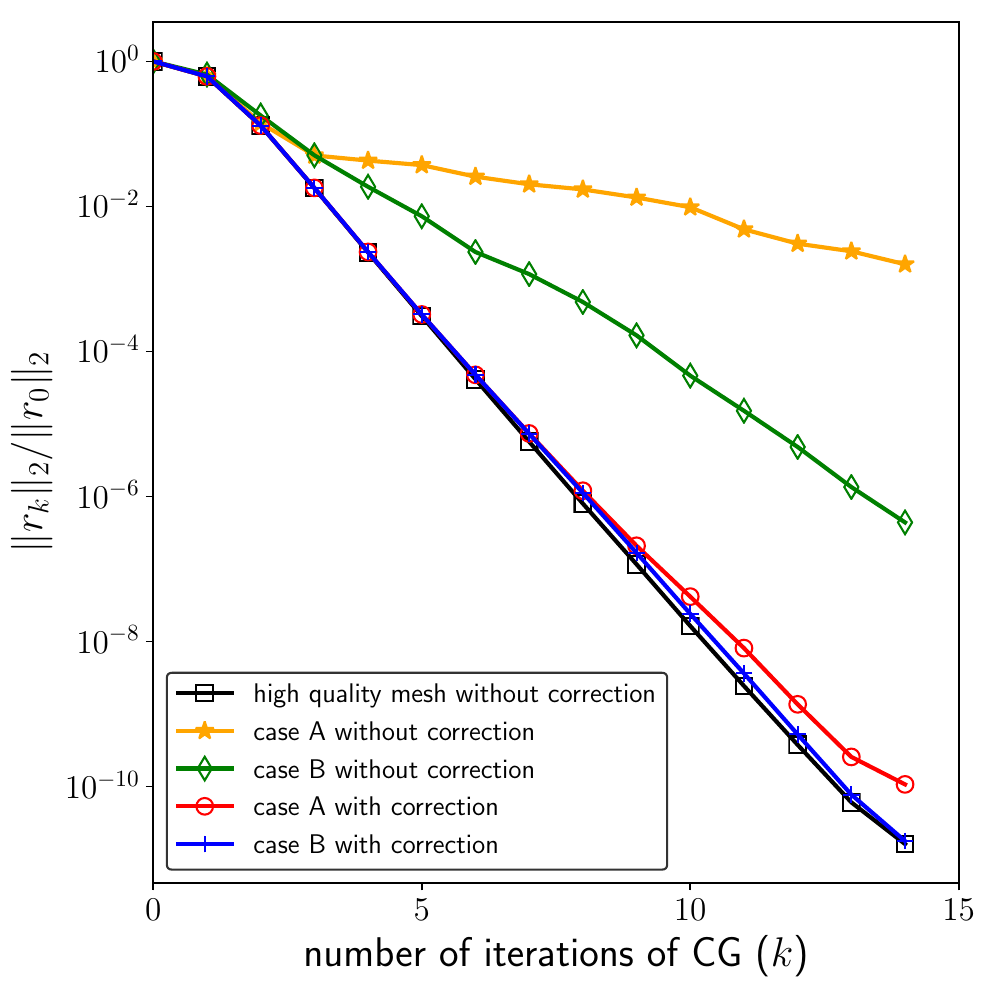}
    \caption{$P_2$ element}
  \end{subfigure}
  \caption{Relative residual after each multigrid (symmetric Gauss--Seidel
  as global smoother) preconditioned CG iteration with and without local
  correction for the linear elasticity problem on the lattice mesh with
  low quality regions on all levels (Case~A) and low quality regions on
  all levels except the finest level (Case~B).}
  \label{fig:res_lattice}
\end{figure}

\begin{figure}
  \centering
  \begin{subfigure}{.5\textwidth}
    \centering
    \includegraphics[width=0.7\linewidth]{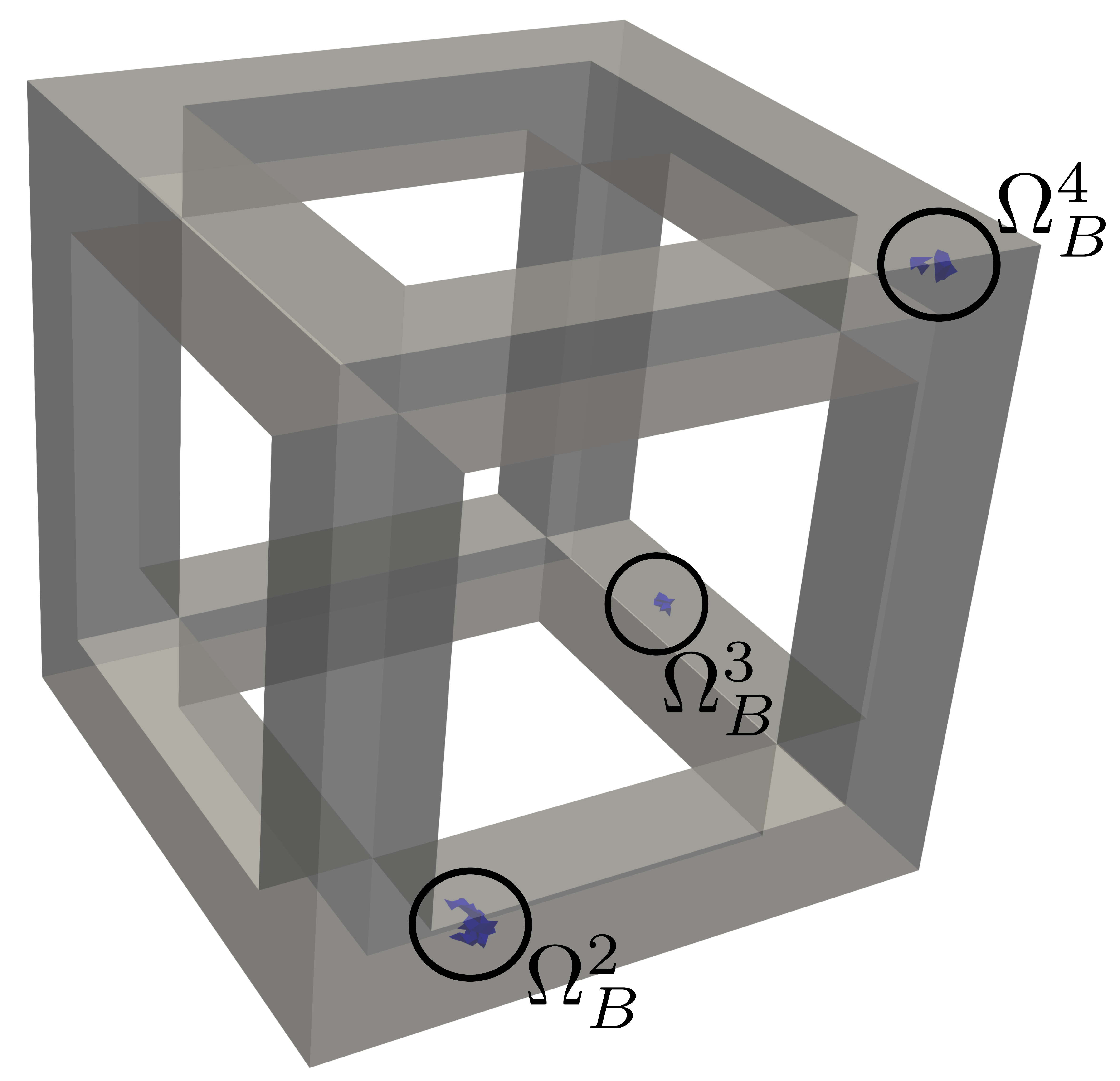}
    \caption{$P_1$ element}
    \label{cap1}
  \end{subfigure}%
  \begin{subfigure}{.5\textwidth}
    \centering
    \includegraphics[width=0.7\linewidth]{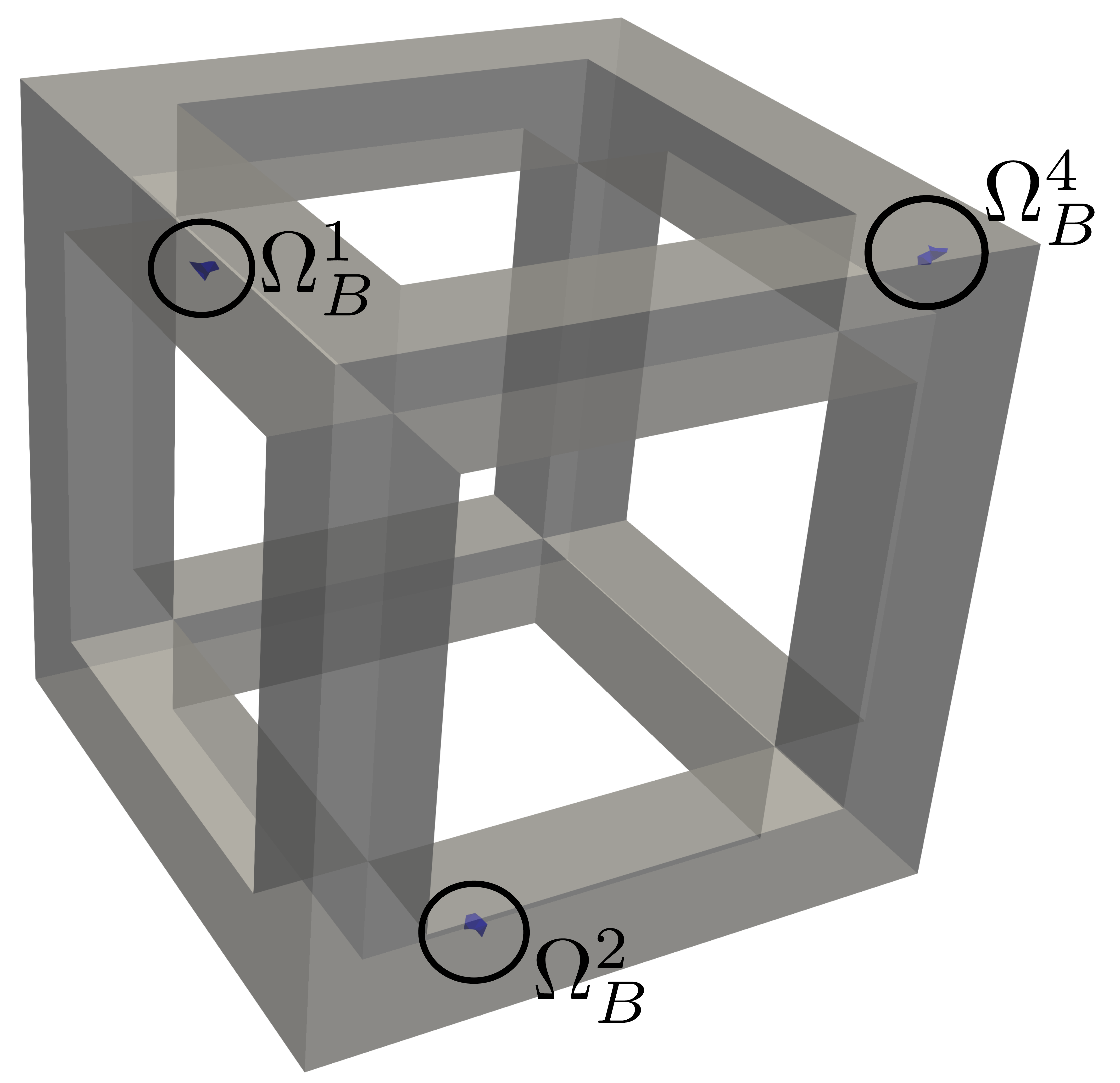}
    \caption{$P_2$ element}
    \label{cap2}
  \end{subfigure}
  \caption{Locations on the finest lattice grid where the residual is
  large after ten iterations of multigrid preconditioned CG.}
  \label{fig:contour_lattice}
\end{figure}

With symmetric Gauss--Seidel applied as the global smoother, the
computed residual after each CG iteration, for $P_1$ and $P_2$ elements,
is shown in \cref{fig:res_lattice} for Case~A and Case~B.
The convergence rate with low quality meshes at all levels (Case~A) is
poor compared to the reference case. The performance for Case~B (high
quality fine grid, low quality coarse grids) appears reasonable for
$P_{1}$ elements, but is considerably worse than the reference case for
$P_{2}$ elements. Using the local correction smoother, the reference
rate of convergence is recovered. \Cref{fig:contour_lattice} shows the
locations on the finest grid where, after 10 iterations, the residual is
large for Case~A (absolute value of the residual is greater
than~$10^{-6}$). Comparing to \cref{fig:domain_lattice}, the large
residuals coincide with regions of low cell quality.

\subsubsection{Dumbbell-like structure}

\begin{figure}
  \centering
  \begin{subfigure}{.8\textwidth}
    \centering
    \includegraphics[width=0.5\linewidth]{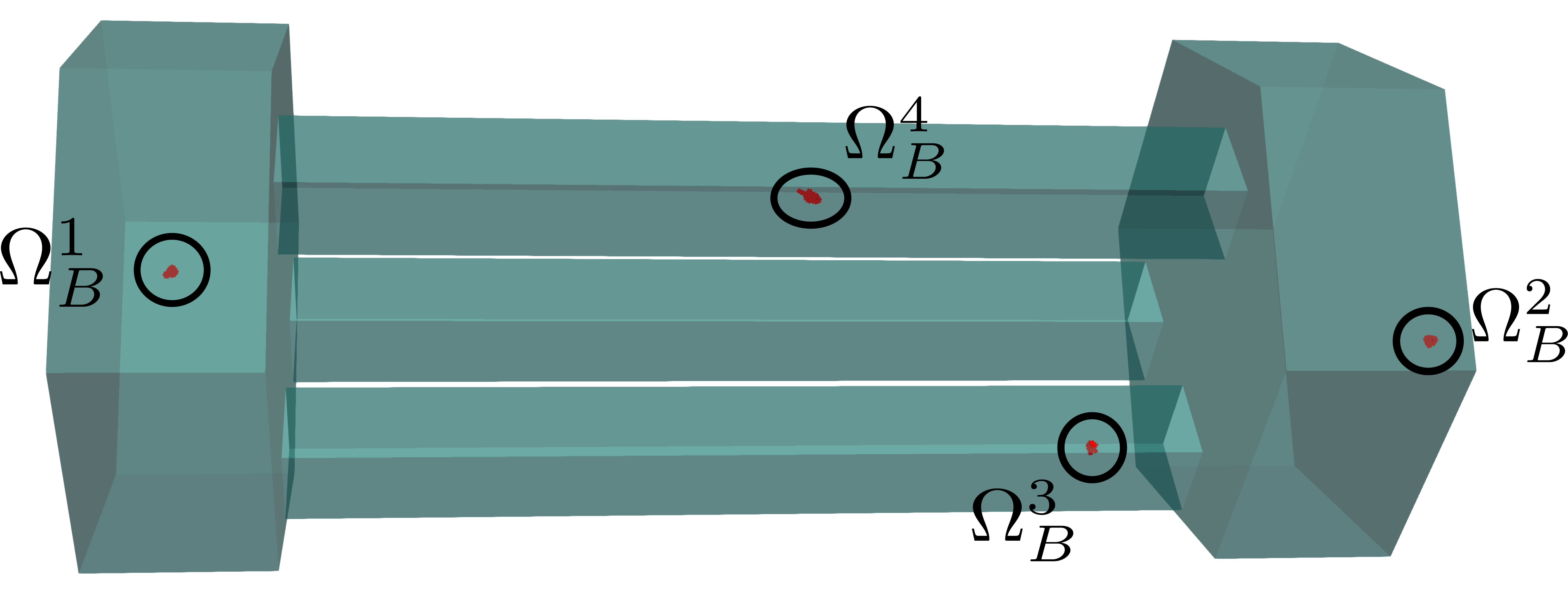}
    \caption{level 1 for $P_1$ element}
  \end{subfigure}%
  \hfill
  \begin{subfigure}{.8\textwidth}
    \centering
    \includegraphics[width=0.5\linewidth]{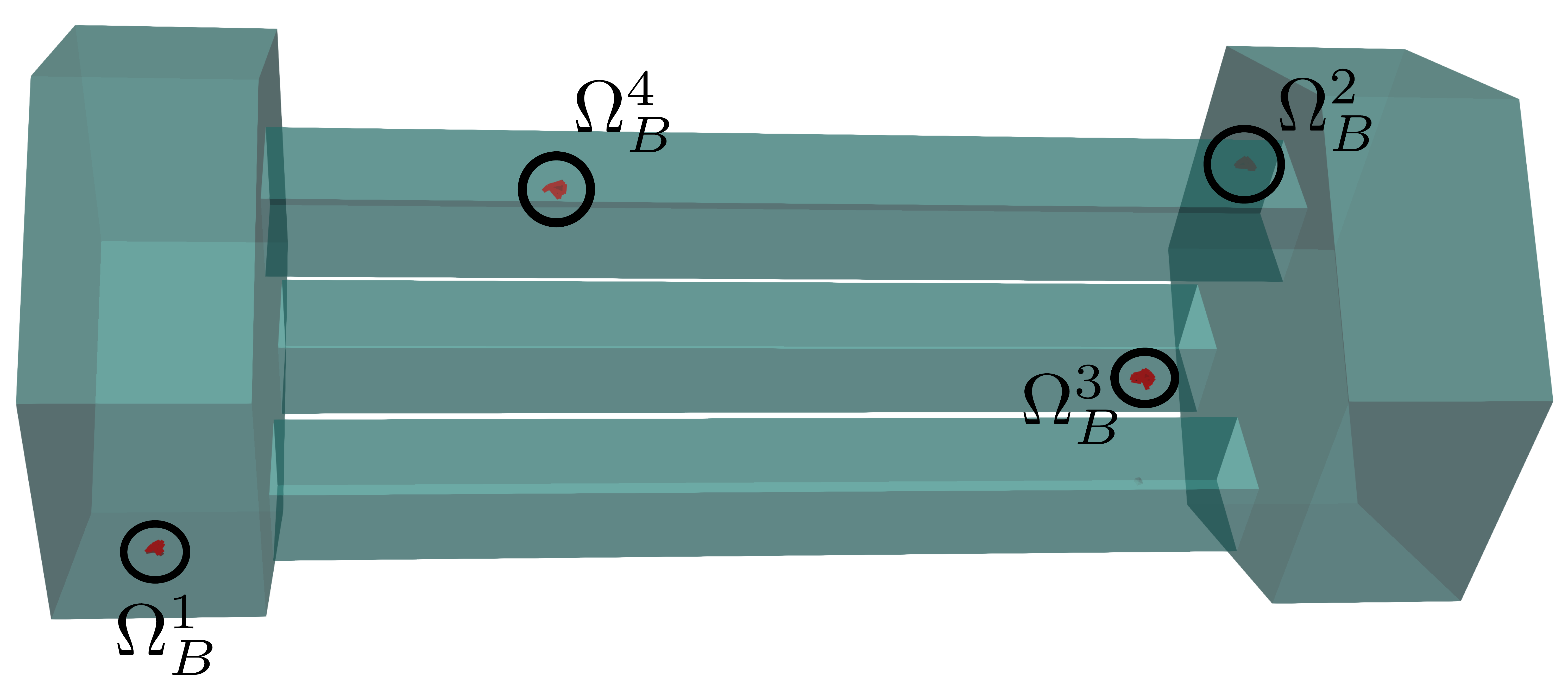}
    \caption{level 1 for $P_2$ element}
  \end{subfigure}
  \caption{A dumbbell-like structure and the positions of low quality
  cells on the finest grid of the $P_1$ and $P_2$ elements.}
  \label{fig:domain_struct}
\end{figure}

Finally we consider a dumbbell-like structure with hexagonal ends
connected by three slender bars, shown in \cref{fig:domain_struct}. We
take $f = (0, 0, -10^3)$ and boundary conditions: $u = (0,0,0)$ on the
left-hand most boundary, $\sigma \cdot n =(10^3,0,0)$ on the right-hand
most boundary, and $\sigma \cdot n = (0,0,0)$ on other parts of
boundary.
Summary data for the meshes is given in \cref{table:size_structure}.
Different meshes are used for the $P_1$ and $P_2$ element cases. Note
that the coarsening rates vary between levels, with more aggressive
coarsening from the simulation (finest) level to the second level.

\begin{table}
  \footnotesize
  \centering
  \begin{tabular}{lccccccc}
  \hline
  \multicolumn{8}{c}{$P_1$ element} \\
  \hline
  level & \multicolumn{1}{c|}{cell size}
  & \multicolumn{1}{c}{$\gamma_{\min}(\Omega_{\text{high}})$}
  & \multicolumn{1}{c|}{$\gamma_{\min}(\Omega_{\text{low}})$}
  & \begin{tabular}[c]{@{}c@{}}number of\\ cells in $\Omega$ \end{tabular}
  & \multicolumn{1}{c|}{\begin{tabular}[c]{@{}c@{}}number of
  \\ cells in $\Omega_B$ \end{tabular}}
  & \begin{tabular}[c]{@{}c@{}}number of \\ DOFs in $\Omega$ \end{tabular}
  & \begin{tabular}[c]{@{}c@{}}number of \\ DOFs in $\Omega_B$ \end{tabular}
  \\
  \hline
  1 (fine) & \multicolumn{1}{c|}{$0.025$}  & $0.150$
  & \multicolumn{1}{c|}{$3.18\times10^{-8}$} & \num{3982354}
  & \multicolumn{1}{c|}{\num{788}}      & \num{2130582}  & \num{876}   \\
  2  & \multicolumn{1}{c|}{$0.08$}  & $0.204$
  & \multicolumn{1}{c|}{$6.94\times10^{-8}$} & \num{139469}
  & \multicolumn{1}{c|}{\num{538}}      & \num{90027}    & \num{699}    \\
  3  & \multicolumn{1}{c|}{$0.2$}   & $0.273$
  & \multicolumn{1}{c|}{$2.93\times10^{-8}$} & \num{12627}
  & \multicolumn{1}{c|}{\num{454}}      & \num{10455}    & \num{615}      \\
  4 (coarse) & \multicolumn{1}{c|}{$0.5$}   & $0.179$
  & \multicolumn{1}{c|}{$8.92\times10^{-8}$}        & \num{1345}
  & \multicolumn{1}{c|}{\num{164}}        & \num{1515}      & \num{216}     \\
  \hline
  \multicolumn{8}{c}{$P_2$ element} \\
  \hline
  level & \multicolumn{1}{c|}{cell size}
  & \multicolumn{1}{c}{$\gamma_{\min}(\Omega_{\text{high}})$}
  & \multicolumn{1}{c|}{$\gamma_{\min}(\Omega_{\text{low}})$}
  & \begin{tabular}[c]{@{}c@{}}number of\\ cells in $\Omega$ \end{tabular}
  & \multicolumn{1}{c|}{\begin{tabular}[c]{@{}c@{}}number of
  \\ cells in $\Omega_B$ \end{tabular}}
  & \begin{tabular}[c]{@{}c@{}}number of \\ DOFs in $\Omega$ \end{tabular}
  & \begin{tabular}[c]{@{}c@{}}number of \\ DOFs in $\Omega_B$ \end{tabular}
  \\
  \hline
  1 (fine)  & \multicolumn{1}{c|}{$0.04$}  &  $0.220$
  & \multicolumn{1}{c|}{$3.26\times10^{-8}$}  & \num{1035068}
  & \multicolumn{1}{c|}{\num{600}}       & \num{4453713}   & \num{3771}   \\
  2   & \multicolumn{1}{c|}{$0.12$}  &  $0.211$
  & \multicolumn{1}{c|}{$6.27\times10^{-8}$}  & \num{48880}
  & \multicolumn{1}{c|}{\num{381}}       & \num{237942}    & \num{2664}    \\
  3  & \multicolumn{1}{c|}{$0.25$}   &  $0.267$
  & \multicolumn{1}{c|}{$8.34\times10^{-8}$}  & \num{5059}
  & \multicolumn{1}{c|}{\num{253}}       & \num{28863}     & \num{1926}     \\
  4 (coarse) & \multicolumn{1}{c|}{$0.5$}    & $0.179$
  & \multicolumn{1}{c|}{$8.92\times10^{-8}$}  & \num{1345}
  & \multicolumn{1}{c|}{\num{164}}         & \num{8532}      & \num{1116}  \\
  \hline
  \end{tabular}
  \caption{Cell quality data (minimum normalized radius ratio $\gamma$)
  and problem size for each level of the dumbbell--like hierarchy
  meshes.}
  \label{table:size_structure}
\end{table}

\begin{figure}
  \centering
  \begin{subfigure}{.5\textwidth}
    \centering
    \includegraphics[width=0.9\linewidth]{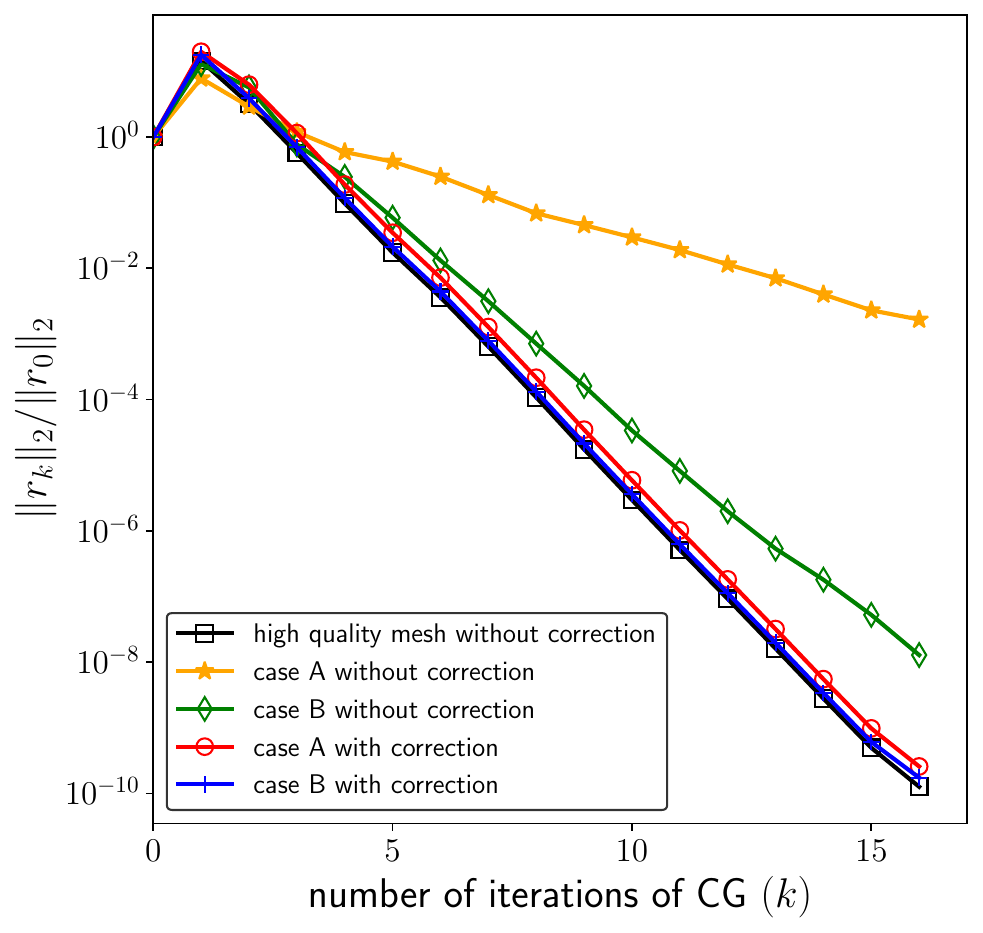}
    \caption{$P_1$ element}
  \end{subfigure}%
  \begin{subfigure}{.5\textwidth}
    \centering
    \includegraphics[width=0.9\linewidth]{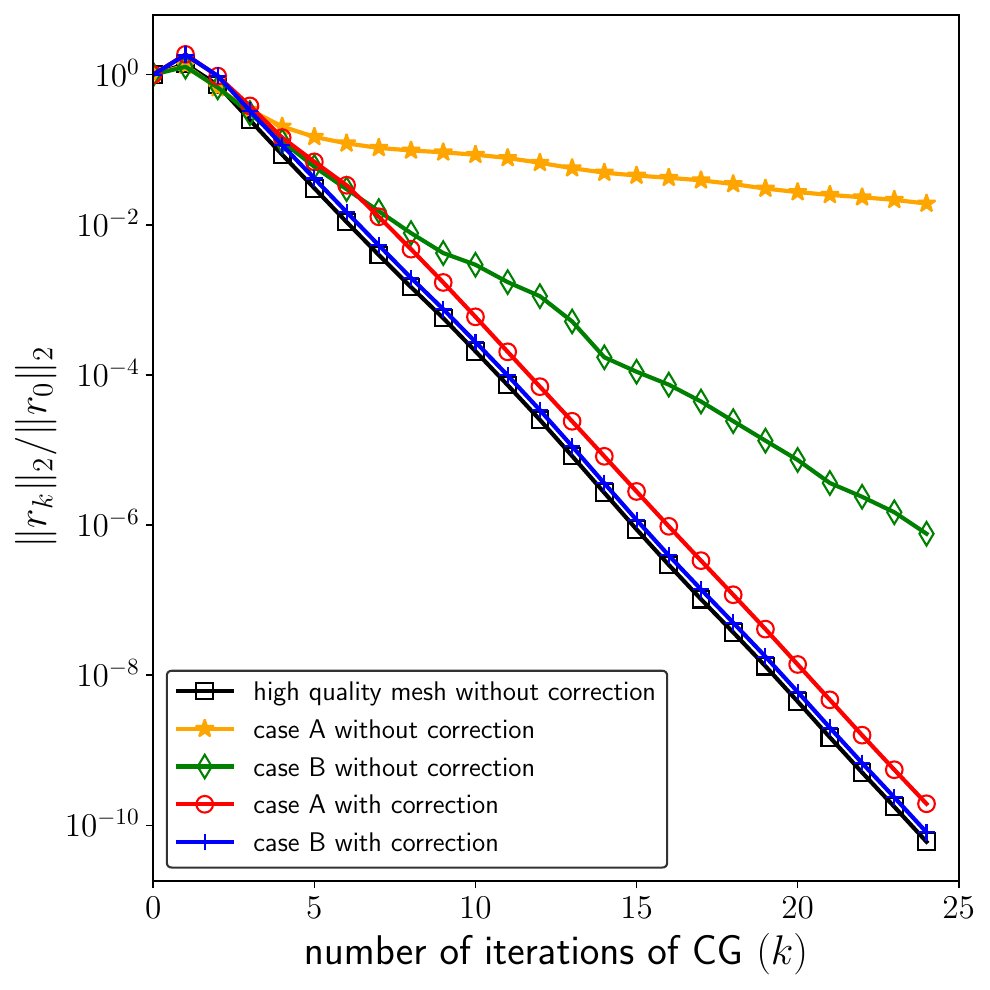}
    \caption{$P_2$ element}
  \end{subfigure}
  \caption{Relative residual after each Krylov iteration for the
  multigrid (Chebyshev global smoother with adjusted largest eigenvalue)
  preconditioned CG solver for solving linear elasticity on the
  dumbbell-like structure. Case~A (low quality regions on all levels)
  and Case~B (low quality regions on levels except the finest level)
  are presented.}
  \label{fig:res_struct}
\end{figure}

We use a Jacobi--preconditioned Chebyshev global smoother for this
example. The smallest eigenvalue in the Chebyshev smoother is set as one
tenth of the largest eigenvalue. The computed residuals at each CG
iteration with the adjusted largest eigenvalue and with and without the
local correction are shown in \cref{fig:res_struct}. Convergence is slow
with low quality meshes and a standard smoother, and particularly so
when the fine grid contains low quality cells (Case~A). As for the
lattice example, for Case~B the $P_{2}$ convergence rate is more heavily
impacted by the low quality cells than the $P_{1}$ case. Applying the
local correction restores the observed convergence close to the
reference case.

The number of CG iterations required to reduce the relative residual to
below $10^{-10}$ are listed in \cref{table:number_struct}, including
with and without the adjusted largest eigenvalues in the Chebyshev
smoother. The convergence rate is improved by using the adjusted maximum
eigenvalue compared to using the largest eigenvalue of the whole system.

\begin{table}
  \centering
  \begin{tabular}{l|cc|cc}
  \hline
  element type & \multicolumn{2}{c|}{$P_1$ element}
  & \multicolumn{2}{c}{$P_2$ element}  \\
  \hline
  reference case   & $17$    &    & $24$     &    \\
  \hline
  case & Case A & Case B & Case A   & Case B \\
  \hline
  low quality mesh without correction    & $> 100$     & $25$   & $\gg 100$  &  $49$  \\
  \hline
  \begin{tabular}[l]{@{}l@{}}low quality mesh with correction\\
  using unadjusted $\lambda_{\max}$ in Chebyshev\end{tabular}
  & $25$    & $22$    & $48$    & $40$    \\
  \hline
  \begin{tabular}[l]{@{}l@{}}low quality mesh with correction\\
  using adjusted $\lambda_{\max}$ in Chebyshev\end{tabular}
  & $18$   & $17$     & $25$    & $24$     \\
  \hline
  \end{tabular}
  \caption{Number of multigrid (Chebyshev global smoother)
  preconditioned CG iterations needed to reduce the relative residual to
  $10^{-10}$ for linear elasticity on the the dumbbell--like structure
  with low quality regions on all levels (Case~A) and low quality
  regions on levels except the finest level (Case~B).}
  \label{table:number_struct}
\end{table}

\section{Conclusions}
\label{sec:conclusion}

Geometric multigrid on non-nested, unstructured grids has been
considered in the presence of a small number of low quality cells, which
can characterise meshes of geometrically complex domains (at the very
least in intermediate grids). It was observed that the performance of
the geometric multigrid method degrades significantly when a mesh
contains a small number of low quality cells, with the poor convergence
attributed to the local failure of smoothers in regions close to the low
quality cells. A global--local combined smoother was developed to
overcome this issue. The smoother involves application of a global
smoother on the entire grid combined with a local correction on
subdomains with low cell quality.

We have demonstrated the proposed smoother on several numerical
examples. It eliminates errors in low cell quality regions that are not
removed by a standard smoother. It was shown that the slow convergence
rate for low quality meshes can be restored to the high quality mesh
reference rate by applying the combined smoother. In particular, if the
fine grid is high quality and only coarse grids have low quality
regions, the finite element discretisation error is barely influenced
and the local correction improves the convergence rate of multigrid.
This work improves the robustness for multigrid on complex geometric
domains and opens up the possibility of the high performance, geometric,
scalable multigrid solvers to solve complicated engineering applications
at a system level.

The non-nested geometric approach can offer an alternative to algebraic
multigrid in some cases. A question is the extension of the approach to
algebraic multigrid (AMG).  Work is currently underway investigating
coarsening strategies, transfer operators and smoothers for AMG that are
robust with respect to cell quality.

\section*{Acknowledgements}

YC was supported by the Youth Program of the Natural Science Foundation
of Jiangsu Province (No.~BK20230466), the Jiangsu Funding Program for
Excellent Postdoctoral Talent (No.~2022ZB584), and Jiangsu Shuangchuang
Project (JSSCTD202209). GNW acknowledges support from the Engineering
and Physical Sciences Research Council under grants EP/V001396/1,
EP/S005072/1, EP/W00755X/1 and EP/W026635/1.

\bibliographystyle{plainnat}
\bibliography{sample}
\end{document}